\documentclass[final,authoryear,5p,times,twocolumn]{elsarticle}
\usepackage{geometry}
\usepackage{lineno}

\usepackage[utf8]{inputenc}
\usepackage{graphicx}
\usepackage{lscape}
\usepackage{rotating}
\usepackage{xfrac}
\usepackage{epsfig}
\usepackage{amssymb}
\usepackage{amsmath}
\usepackage{journals}
\usepackage{wasysym}
\usepackage[justification=justified,singlelinecheck=false]{caption}
\usepackage{color}
\usepackage[colorlinks=True,
          urlcolor=blue,
          linkcolor=blue,
          citecolor= blue,
          pdftex]{hyperref}
\usepackage{cleveref}
\usepackage{longtable}
\usepackage[usenames,dvipsnames]{xcolor}
\usepackage{bigints}
\usepackage{soul,color}
\usepackage{color, colortbl}

\newcommand{\nad}[1]{#1}        

\newcommand{\Jupiter}{BG-G14}
\newcommand{\RD}{\mbox{$R_{50}$}}
\newcommand{\mis}{\mbox{$M$}}
\newcommand{\figref}[1]{Fig.~\ref{#1}}
\newcommand{\Figref}[1]{Figure~\ref{#1}}

\newcommand{\eqnref}[1]{Eq.~\ref{#1}}

\newcommand{\tabref}[1]{Tab.~\ref{#1}}
\newcommand{\Tabref}[1]{Table~\ref{#1}}

\newcommand{\secref}[1]{Sec.~\ref{#1}}

\newcommand{\mdot}{\;\;{\textrm.}}
\newcommand{\mcom}{\;\;{\textrm,}}
\newcommand{\bad}{\mbox{$b_{10}$}}
\newcommand{\bed}{\mbox{$b_{11}$}}
\newcommand{\bquad}{\mbox{$b_{2}$}}
\newcommand{\bocto}{\mbox{$b_{3}$}}

\begin{document}


\begin{frontmatter}

\title{Physical conditions for Jupiter-like dynamo models}
\author{Lúcia D. V. Duarte\fnref{label1}}
\ead{lduarte@astro.ex.ac.uk}
\author{Johannes Wicht\fnref{label2}}
\author{Thomas Gastine\fnref{label3}}

\fntext[label1]{College of Engineering, Mathematics and Physical Sciences, University of Exeter, Physics building, Stocker Road, Exeter, United Kingdom, EX4 4QL}
\fntext[label2]{Max-Planck-Institut für Sonnensystemforschung, Justus-von-Liebig-Weg 3, 37077 Göttingen, Germany}
\fntext[label3]{Institut de Physique du Globe de Paris, Sorbonne Paris Cit\'e,
Universit\'e Paris-Diderot, UMR 7154 CNRS, 1 rue Jussieu, F-75005 Paris,
France}
\address{}

\begin{abstract}

The Juno mission will measure Jupiter's magnetic field with unprecedented
precision and provide a wealth of additional data that will allow us to
constrain the planet's interior structure and dynamics.
Here we analyse $66$ different numerical simulations in order to explore the sensitivity
of the dynamo-generated magnetic field to the planets interior properties.
Jupiter field models based on pre-Juno data and up-to-date interior models based on
\textit{ab initio} simulations serve as benchmarks.
Our results suggest that Jupiter-like magnetic fields can be found
for a number of different models.
These complement the steep density gradients in the outer part of
the simulated shell with an electrical conductivity profile that mimics
the low conductivity in the molecular hydrogen layer and thus renders
the dynamo action in this region largely unimportant.
We find that whether we assume an ideal gas or use the more realistic interior model
based on \textit{ab initio} simulations makes no difference.
However, two other factors are important. A low Rayleigh number leads to
a too strong axial dipole contribution while the axial dipole dominance is lost
altogether when the convective driving is too strong. The required intermediate range that
yields Jupiter-like magnetic fields depends on the other system properties.
The second important factor is the convective magnetic Reynolds number radial profile $Rm_c(r)$, basically a product of the non-axisymmetric flow velocity and electrical conductivity.
We find that the depth where $Rm_c$ exceeds about $50$ is a good proxy for the top of the
dynamo region. When the dynamo region sits too deep, the axial dipole is once more
too dominant due to geometric reasons.
Extrapolating our results to Jupiter and the result suggests that the Jovian dynamo extends to $95$\% of the planetary radius.

The zonal flow system in our simulations is dominated by an equatorial jet which remains
largely confined to the molecular layer.
Where the jet reaches down to higher electrical conductivities, however, it gives rise
to a secondary $\alpha\Omega$ dynamo that modifies the dipole-dominated field
produced deeper in the planet. This secondary dynamo can lead to strong magnetic field patches at lower latitudes that seem compatible with the pre-Juno field models.

\end{abstract}

\begin{keyword}

Atmospheres, dynamics
\sep Jupiter, interior
\sep Variable electrical conductivity
\sep Numerical dynamos

\end{keyword}

\end{frontmatter}

\section{Introduction}
\label{intro}

The interior dynamics of Jupiter has been the topic of an increasing number of
studies over the last ten years
\citep{Heimpel05,Lian08,Stanley09,Kaspi09,Heimpel11,Gastine12,
Duarte13,Gastine14,Jones14,Heimpel16}.
The growing interest is at least partially motivated by two Jovian space missions.
NASA's Juno spacecraft arrived in summer 2016 and  started to measure the
planet's magnetic field with unprecedented precision. It will also provide important
information on the inner structure and dynamics, for example via gravity data.
ESA's Jupiter system mission Juice is scheduled to be launched in 2022.

Recent models for Jupiter's interior structure combine pre-Juno gravity
and planetary figure measurements with refined equations of state that
are based on \textit{ab initio} calculations \citep{French12,Nettelmann12}.
These models assume a small rocky core of uncertain size and a two layer
Hydrogen and Helium envelope where the inner layer contains more
heavy elements than the outer.
When pressures are high enough at about $80$ to $90$\% of Jupiter's radius,
hydrogen undergoes a phase transition from the molecular to a metallic state \citep[e.g.][]{Chabrier92,Fortney10,Nettelmann12}.
Since this transition lies beyond the triple point \citep{French12}
there is no sharp change in the physical properties.
Using advanced \textit{ab initio} calculations, \cite{French12} shows that
the electrical conductivity, a particularly important property for the dynamo process,
rises steeply with depth at a super-exponential rate in the molecular layer and then
more smoothly transitions into the metallic region
where the gradient becomes much shallower (see \figref{fig:vconds}).

Magnetic field models in the pre-Juno era rely on a few flybys and sometimes
auroral information to constrain spherical harmonic
surface field contributions up to degree $\ell\!=\!4$
\citep{Connerney98,Grodent08,Hess11}
or $\ell\!=\!7$ at best \citep{Ridley12,Ridley16}.
Due to its dedicated polar orbit, Juno is expected to constrain
models up to $\ell\!=\!15$ or higher. This exceeds the resolution
available for Earth where the crustal field shields harmonics beyond $\ell\simeq 14$.

While the magnetic field offers indirect clues for the deeper processes
the surface dynamics can be inferred more directly,
for example by tracking cloud features.
The surface winds are dominated by a system of zonal jets where
a fast prograde equatorial jet is flanked by several additional jets of
alternating retrograde and prograde direction.
At least the equatorial jet could be a geostrophic
structure that reaches through the planets and is maintained
by Reynolds stresses, a statistical correlation between smaller-scale
convective flow components
\citep{Christensen01,Heimpel05,Gastine12,Gastine12a}.
This is less clear for the flanking jets which may be
much shallower thermal-wind driven structures \citep{Kaspi09}.
Constraining the depth of Jupiter's jet system is one of the main
objectives of the Juno mission.

Though the \textit{ab initio} simulations may not support a clear separation,
traditional simulations of Jupiter's internal dynamics concentrated on either describing
the deeper dynamo thought to operate in the metallic hydrogen layer
or on the jet dynamics in the molecular envelope.
While the latter are very successful in describing the observed zonal jet
structure \citep{Heimpel05,Gastine14a,Heimpel16} the dynamo simulation have proven
to be more problematic.
Since Jupiter's magnetic field has a very Earth-like configuration is it tempting
to assume that numerical geodynamo simulations capture the dynamics of the
metallic layer. However, geodynamo models typically
neglect compressibility and assume a constant
adiabatic temperature profile in the so-called Boussinesq approximation. Moreover,
the electrical conductivity is constant and rigid flow boundary conditions
are often used that significantly inhibit zonal winds \citep{Olson99,Christensen07}.
These simulations show that dipolar and thus Earth-like or Jupiter-like magnetic fields
can only be expected when the system is not driven too strongly, i.e.~the Rayleigh number
remains in a range where inertial effects are small \citep{Christensen06}.

More recent simulations have shown that it becomes increasingly complicated to
maintain dipole-dominated fields when modifying the models to better represent
gas planets.
Using stress free rather than rigid flow boundary conditions allows strong
Reynolds-stress driven zonal winds to develop, which are always highly geostrophic
and thus reach through the whole gaseous envelope.
The competition between these winds and strong dipolar fields
plays an important role in determining whether the magnetic field becomes axial dipole-dominated
or multipolar, i.e.~more complex without a dominant axial dipole
contribution \citep{Grote00,Busse06,Simitev09,Sasaki11,Schrinner12,Gastine12a}.
Zonal flows tend to promote weaker multipolar fields while strong dipole fields can
suppress the zonal flows via Lorentz forces. Dipole-dominated dynamos
thus require a certain balance between flow vigour and dipole field amplitude.

A consequence of this competition is the bistability found at not too
large Rayleigh numbers where dipole and multipole solutions coexist
at identical parameters \citep{Gastine12a}. The multipolar branch is reached when starting
a simulation with a weak field and is characterized by stronger zonal flows.
Establishing a solution on the dipolar branch, on the other hand,
requires to start with a strong dipole that sufficiently
suppresses the zonal flows \citep[e.g.][]{Schrinner12}.
When the Rayleigh number is increased beyond a certain point only
the multipolar branch remains. However, the simple rule that describes
this transition for Earth-like dynamos in terms
of the relative importance of inertia \citep{Christensen06}
does mostly not apply in gas giants \citep{Duarte13}.

Including Jupiter-like density profiles in the so-called anelastic approximation
\citep{Glatz123,Glatz84,Brag95,Lantz99} yields further difficulties.
The density stratification leads to convective flows where the
amplitude increases with radius while the length scale decreases.
The zonal flow system changes less dramatically but nevertheless significantly:
the equatorial jet becomes somewhat more confined and increases in amplitude while
the mid to higher latitude jets slow down \citep{Gastine12}.

More successful are integrated models that include the molecular
envelope and more specifically the steep decrease in electrical
conductivity \citep{Gastine14,Jones14}.
This allows the strong equatorial jet to remain mostly
constrained to the weakly conducting outer envelope and thus
participate little in the primary dynamo action \citep{Gastine12a,Gastine14}.
The dipole field actually contributes to establishing itself by pushing the
equatorial jet to the weakly conducting shell and by braking the remaining
high to mid latitude zonal flows via Lorentz forces \citep{Duarte13}.
When the weakly conducting shell is too thick, however, the region where the Lorentz force
can counteract zonal wind production via Reynolds stresses becomes too restricted and
the dynamo ends up producing a multipolar field.

Another effect that can help establishing a dipole-dominated field
is an increased magnetic Prandtl number $Pm\!=\!\nu/\lambda$ where $\nu$ is the kinematic viscosity and $\lambda\!=\!1/(\sigma\mu_0)$
the magnetic diffusivity \citep
{Duarte13,Schrinner14,Jones14,Raynaud15}.
Increasing $Pm$ is equivalent to increasing the electrical conductivity $\sigma$ which leads to
a more efficient dynamo and likely also stronger Lorentz forces that can more easily
balance zonal flows.
Several authors also report that either decreasing or increasing the Prandtl number $Pr\!=\!\nu/\kappa$ (ratio of kinematic
viscosity $\nu$ to thermal diffusivity $\kappa$), from a typical value
of $Pr\!=\!1$ used in many simulations to $0.1$ or $10$, respectively, may also help
\citep{Jones14,Yadav15a,Yadav15b}.
\cite{Jones14} argues that at low $Pr$ the convection is more evenly distributed throughout
the shell which leads to a less dominant equatorial jet
and stronger dipole field generated at depth, while higher $Pr$ means reduced inertia \citep{Yadav15a}.
The effect is potentially important for Jupiter where $Pr$ may be as low as $10^{-2}$
at depth \citep{French12}.

The Ekman number $E$ is another parameter that can influence the magnetic field configuration.
$E$ is a measure for the ratio of viscous to Coriolis forces in the flow force balance.
Because of the small viscosity and fast planetary rotation, Jupiter's Ekman number is only about
$10^{-18}$. For the simulations, however, a much higher viscosity
is assumed to damp the small scale convection that cannot be resolved numerically
and $E$ is typically of order $10^{-4}$ or $10^{-5}$.
Boussinesq dynamo simulations suggest that a lower $E$ promotes dipolar fields
because the stronger Coriolis forces help to organize large scale magnetic field generation
\citep{Christensen06,Wicht10}.
Since \citet{Heimpel11,Duarte13} suggest that a lower $10^{-5}$ may also help to establish
dipolar dynamos in anelastic simulations it
seems  important to further explore this issue.

Many authors drive convection in their Jupiter models from the bottom
\citep{Gastine12a,Duarte13,Gastine14} as would be more appropriate for Earth.
Heat enters the modelled spherical shell through the inner and leaves it through the outer boundary.
However, internal heat sources seem more realistic for modelling the secular
cooling that drives convection in Jupiter. \cite{Jones14} reports that
internal heating makes it less likely to find a dipole-dominated dynamo.

In the most realistic simulation to date, \cite{Gastine14} reproduce many distinct features of
the pre-Juno Jovian magnetic field.
This includes the field strength, dipole dominance, dipole tilt, magnetic power spectrum and secular variation
estimates \citep{Connerney98,Ridley16}. Their model, that we will refer to as G14 in the following,
covers $99$\% of the Jovian density profile suggested by \cite{French12} and uses an
electrical conductivity profile with a significant conductivity decrease in the molecular layer.
Here we extend the G14 study
by analysing a larger number of different model set-ups with respect to their capability of
reproducing Jupiter's magnetic field. Our data set includes $66$ simulations with different
parameters, different density profiles, different electrical conductivity profiles,
and different driving modes.
The paper is organized as follows: after introducing the numerical model in
Sec.~\ref{model} we analyse the simulation results in Sec.~\ref{results},
and close with a discussion in Sec.~\ref{conclusions}.

\section{Model}
\label{model}

We adopt the anelastic formulation suggested by \cite{Glatz123}, \cite{Brag95} and \cite{Lantz99}.
This allows to consider background density and temperature variations but filters out sounds
waves which would required a significantly smaller time step.
MagIC actually solves for small variations around a background state which is assumed to be hydrostatic,
adiabatic and non-magnetic. In the following, all quantities with a tilde characterize the dimensionless background state.

\subsection{Background state}
\label{background}

\begin{figure}[h!]
\begin{center}
{\centering
      \includegraphics[trim=0.2cm 0.8cm 0.4cm 0cm, width=3.36in]{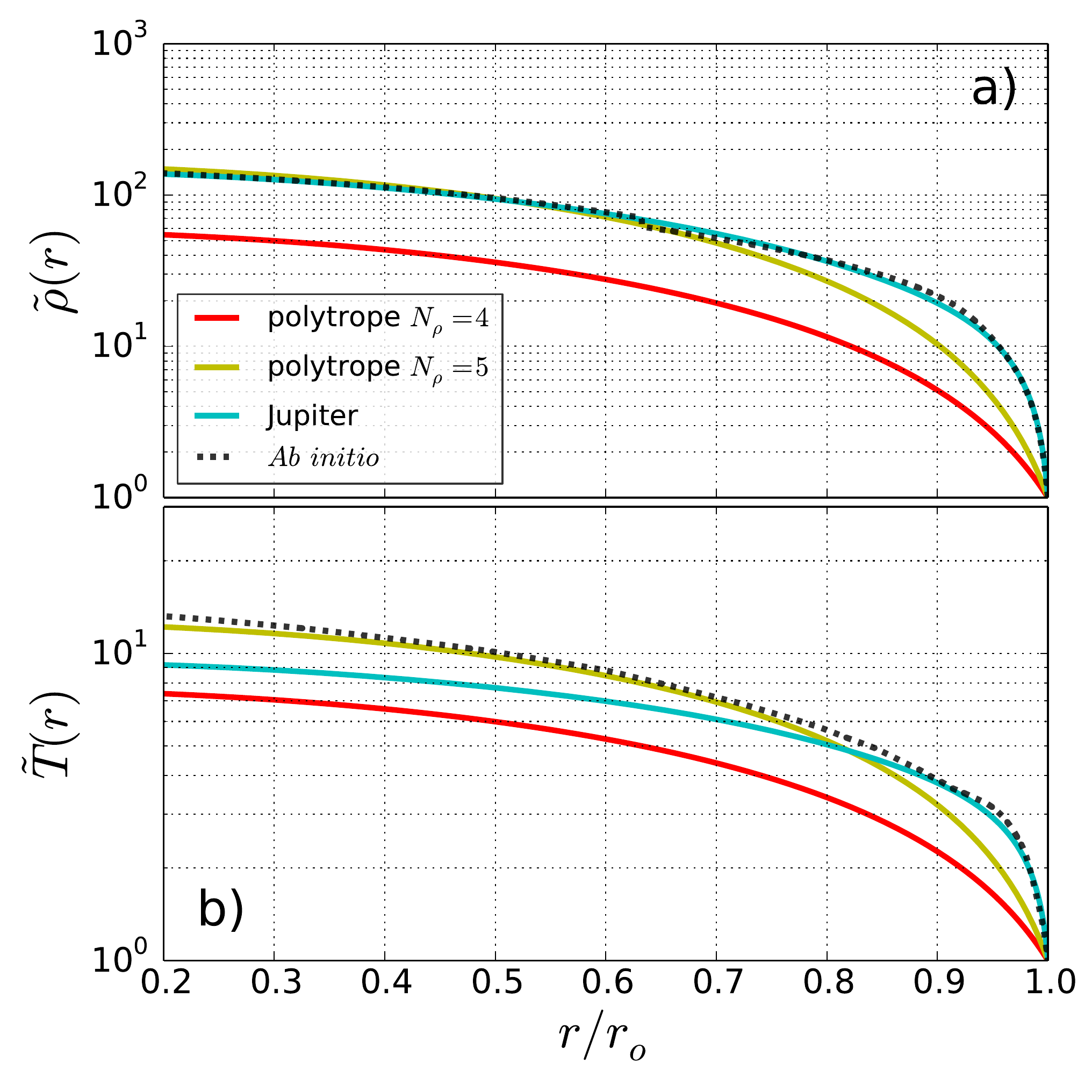}
}
\caption{\small Density (a) and temperature (b) background profiles used in this work. The cyan line corresponds to our fit to the \textit{ab initio} calculations by \cite{Nettelmann12} for the innermost $99\%$ of Jupiter, represented by black dotted line \citep[see also][]{French12}. The overlap is nearly perfect for $\tilde{\rho}$ but the $\tilde{T}$ fit deviates at $r\!<\!0.85r_o$. The red and yellow lines show the polytropic background reference states of $N_\rho\!=\!4$ and $N_\rho\!=\!5$ respectively with polytropic index $n\!=\!2$. To ease the comparison, the density and temperature profiles from the \textit{ab initio} calculation have been normalised by their values at $r\!=\!0.99\,R_{\jupiter}$.
\label{fig:densprofs}}
\end{center}
\end{figure}

\Figref{fig:densprofs}a shows the three different density background profiles considered here.
Several of our models\nad{, corresponding to the cyan profiles in \figref{fig:densprofs}a,b,} rely on the interior properties suggested by \cite{French12}.
Starting point is the \cite{French12} density $\tilde{\rho}(r)$ profile that
is approximated by a polynomial of degree seven.
Due to numerical limitations we can only simulate $99\%$ of Jupiter's radius and
have to ignore the outermost $1$\% where the
density gradient is the steepest.
The density polynomial is then fitted with a power law $\tilde{T}\!=\!\textrm{const. }\tilde{\rho}^{1/n^*}$ \nad{(\figref{fig:densprofs}b)}. We find that an exponent $n^*\!=\!2.22$ yields the best fit. The pressure closely obeys a profile with polytropic index $n\!=\!1$ \citep{Hubbard75}:
\begin{equation}
      \tilde{p} \sim \tilde{\rho}^{(n+1)/n} \sim \tilde{\rho}^2
      \textrm{,}
\end{equation}
where $p$ is pressure.
The difference between the exponents $n$ and $n^*$ demonstrates that Jupiter's interior strongly deviates from an ideal gas behavior where both would be identical ($n\!=\!n^*$).
The product of thermal expansivity and gravity that appears in the buoyancy term of the
Navier-Stokes equation, \eqnref{eq:navierstokeseq}, is directly related to the background temperature
profile via
\begin{equation}
      \frac{1}{\tilde{T}} \frac{\mathrm{d}\tilde{T}}{\mathrm{d} r}=-Di\,\tilde{\alpha}\tilde{g}\mcom
\end{equation}
where
$Di\!=\!\alpha_o g_o d/c_p$ is the dissipation number based on the outer boundary reference gravity $g_o$ and the reference thermal expansivity $\alpha_o$. Here $\tilde{\alpha}\!=\!\alpha/\alpha_o$ is
the normalized thermal expansivity profile, $\tilde{g}\!=\!g/g_o$ the normalized gravity profile, $c_p$ is the heat capacity at constant pressure and $d\!=\!r_o-r_i$ is the difference between outer shell radius $r_o$ and inner shell radius $r_i$ that we use as length scale to non-dimensionalize the equations.
Since this background model has first been adopted
by G14 we refer to this background model as \Jupiter\  in the following.

\nad{The yellow and red profiles in \figref{fig:densprofs} illustrate the two other background
state models used here. For simplicity, we assume a
temperature gradient proportional to radius,
$\mathrm{d}\tilde{T}/\mathrm{d}r\!=\!-Di\,r/r_o$,
and a polytrope-like density profile
$\tilde{\rho}\!=\!\tilde{T}^{1/n}$ with a polytrope index $n\!=\!2$
halfway between $n\!=\!3/2$ for a mono-atomic and $n\!=\!5/2$ for a bi-atomic gas.
For an ideal gas with $\alpha\!=\!1/\tilde{T}$ the temerature gradient
would imply $g\!\sim\!r$ which is realistic for a homogeneous density but not
the polytrope profile.}

\nad{While not realistic for Jupiter these setups nevertheless serve to explore
the sensitivity of our models to the background state.}
The absolute density contrast is controlled by $N_\rho\!=\!\ln\left[\tilde{\rho}(r_i)/\tilde{\rho}(r_o)\right]$,
the number of density scale heights covered between the inner ($r_i$) and the outer ($r_o$)
boundaries. We consider simulations with $N_\rho\!=\!4$ or $N_\rho\!=\!5$ that
we refer to as 'polytrope' models in the following. Figure~\ref{fig:densprofs}a illustrates that the respective density profiles are more gradual than the
\Jupiter\ counterpart where most of the density drop is concentrated at larger radii.

\begin{figure}[h!]
\begin{center}
{\centering
      \includegraphics[trim=0cm 0.5cm -0.2cm 0cm, height=2.55in]{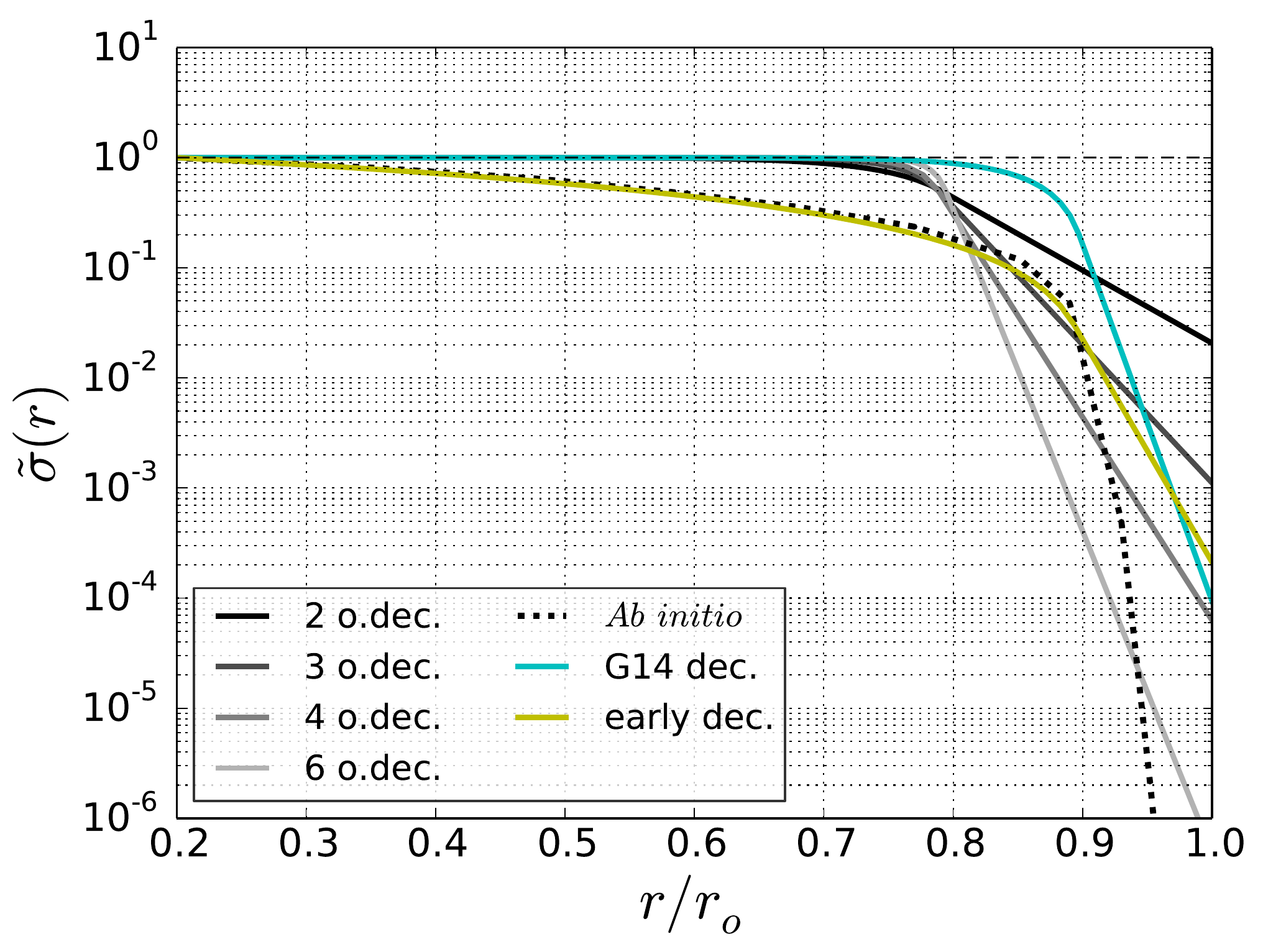}
      }
\caption{\small Radial profiles of normalized electrical conductivity, to illustrate the flexibility of \eqnref{eq:varcondeq}.
The black dotted line corresponds to the \textit{ab initio} calculation from
\cite{French12}. The grey-coloured profiles differ only in $a$ and represent profiles already used in
\cite{Duarte13} with $\sigma_m\!=\!0.5$, $r_m\!=\!80$, and
$a\!=\!9$ (black, approximately 2 orders of magnitude decay), $a\!=\!17$ (dark grey, $\sim3$ oom),
$a\!=\!25$ (medium grey, $\sim4$ oom) and $a\!=\!40$ (light grey, $\sim6$ oom).
The yellow line illustrates the 'earlier decaying'
profile that mimics the \textit{ab initio}
result in the metallic interior, using $a\!=\!1.0$, $\sigma_m\!=\!0.03$, $r_m\!=\!90$.
The cyan line shows the profile
with $a\!=\!13.0$, $\sigma_m\!=\!0.2$, $r_m\!=\!90$ mostly used by G14 and in most of our new simulations.
All profiles have been normalized with the electrical conductivity at the
inner boundary.
\label{fig:vconds}}
\end{center}
\end{figure}

\textit{Ab initio} simulations by \citep{French12} yield an electrical conductivity profile
for the interior of Jupiter that is illustrated in \figref{fig:vconds}). A super-exponential increase in the molecular layer until about $0.9R_{\jupiter}$  smoothly
transitions into a much shallower gradient in the metallic layer.
Since the super-exponential increase causes numerical difficulties, we use several simplified conductivity
profiles with a constant interior conductivity branch that is matched to an exponentially decaying outer branch
via a polynomial that assures a continuous first derivative \citep{Gomez10}:
\begin{equation}
      \tilde{\sigma}(r) =
      \left\{
      \begin{array}{ll}
            1+(\tilde{\sigma}_m-1)\,\Bigg(\dfrac{r-r_i}{r_m-r_i}\Bigg)^a & r<r_m \\
            \tilde{\sigma}_m
            \exp{}\left[a\,\frac{r-r_m}{r_m-r_i}\,
            \frac{\tilde{\sigma}_m-1}{\tilde{\sigma}_m}\right] & r\ge r_m
      \end{array} \right.
       \textrm{.}
\label{eq:varcondeq}
\end{equation}
The tilde once more signifies the non-dimensional background state where we
have used the inner-boundary conductivity value as a reference value $\sigma_i$.
Free parameters are the rate of the exponential decay $a$ and the radius $r_m$
and conductivity value $\sigma_m$ for the transition between both branches.
For convenience we also define the relative transition
radius in percentage: $\chi_m\!=\!r_m/r_o$.
The non-dimensional magnetic diffusivity profile is given by
$\tilde{\lambda}(r)=\lambda(r)/\lambda_i=\sigma_i/\sigma(r) = 1/\tilde{\sigma}(r)$ where
the tilde signifies the non-dimensional background state.

\Figref{fig:vconds} compares the electrical conductivity profiles mostly used
in this study with the profile by \cite{French12}. A series of profiles with
$\chi_m\!=\!0.8$, $\sigma_m\!=\!0.5$ and increasing decay rates from $a\!=\!9$ to $a\!=\!40$ serves to
explore the potential influence of the decay rate.
In the most extreme model with $a\!=\!40$ the conductivity decrease by seven orders of magnitude.
G14 use $\chi_m\!=\!0.9$, $\sigma_m\!=\!0.2$, and $a\!=\!13$ which yields a profile (cyan) where the higher
conductivity inner region reaches further out, the transition to the exponential decay is smoother,
and the total decrease amounts to four orders of magnitude.
That is also the profile adopted for most of our new simulations.
An additional 'early decaying' profile (yellow line in Fig.~\ref{fig:vconds})
tries to model the slower conductivity decrease predicted for the inner
metallic layer with a linear decay rate.

\subsection{Anelastic equations}
\label{anelasticeqs}

We solve for convection and magnetic field generation in a rotating
spherical shell with outer radius $r_o$ and inner radius $r_i$.
The set of anelastic equations that describes the evolution of the
dimensionless velocity $\mathbf{u}$, magnetic field $\mathbf{B}$
and specific entropy $s$ are the Navier-Stokes equation,
dynamo equation, energy equation and continuity equations:
\begin{equation}
\begin{split}
      E\,\bigg(\frac{\partial \mathbf{u}}{\partial t} + \mathbf{u}\cdot\nabla \mathbf{u}\bigg)
      = &- \nabla\frac{p}{\tilde{\rho}} - 2\mathbf{e}_z\times\mathbf{u}
      - \frac{Ra\,E}{Pr\,Di} \frac{d\tilde{T}}{dr} s\,\mathbf{e}_r
\\
      &+ \frac{1}{Pm_i\,\tilde{\rho}}(\nabla\times\mathbf{B})\times\mathbf{B}
      + \frac{E}{\tilde{\rho}}\nabla\cdot\textsf{S} \mcom
\end{split}
\label{eq:navierstokeseq}
\end{equation}
\begin{equation}
      \frac{\partial \mathbf{B}}{\partial t} = \nabla\times(\mathbf{u}\times\mathbf{B})
      - \frac{1}{Pm_i}\nabla\times(\tilde{\lambda}\nabla\times\mathbf{B}) \mcom
\label{eq:inductioneq}
\end{equation}
\begin{equation}
\begin{split}
      \tilde{\rho}\,\tilde{T}\,\bigg(\frac{\partial s}{\partial t} + \mathbf{u}\cdot\nabla s\bigg)
            = &\frac{1}{Pr}\nabla\cdot (\tilde{\rho}\tilde{T}\nabla s)
                  + \epsilon\tilde{\rho}
\\
      &+ \frac{Pr\,Di}{Ra} \left[ \textsf{S}^2
            + \frac{\tilde{\lambda}}{Pm_i^2\,E} (\nabla\times\mathbf{B})^2 \right]
      \mcom
\end{split}
\label{eq:energyeq}
\end{equation}
\begin{equation}
      \nabla\cdot (\tilde{\rho}\mathbf{u}) = 0 \mcom
\label{eq:divfloweq}
\end{equation}
\begin{equation}
      \nabla\cdot\mathbf{B} = 0 \mdot
\label{eq:divfieldeq}
\end{equation}
Here $\textsf{S}$ is the traceless rate-of-strain tensor for an homogeneous kinematic viscosity,
\begin{equation}
      \textsf{S} = 2\tilde{\rho}\bigg[\textsf{e}_{ij}-\frac{1}{3}\delta_{ij}\nabla\cdot\mathbf{u}\bigg] \;\;\;\textrm{  and   }\;\;\;
      \textsf{e}_{ij} = \frac{1}{2}\bigg(\frac{\partial u_i}{\partial x_j}+\frac{\partial u_j}{\partial x_i}\bigg) \mcom
\label{eq:straintensor}
\end{equation}
where $\delta_{ij}$ is the identity matrix. The terms inside the square brackets of Eq.~\ref{eq:energyeq} correspond to the viscous and ohmic heating contributions.

These equations have been non-dimensionalized using
shell thickness $d\!=\!r_o-r_i$ as the length scale and the viscous
diffusion time $\tau_{\nu}\!=\!d^2/{\nu}$ as a timescale.
Temperature and density are non-dimensionalized by their values at
the outer boundary, $T_o$ and $\rho_o$. We employ either constant
entropy or constant entropy flux boundary conditions. In the former case,
the imposed super-adiabatic contrast $\Delta s$ across the
shell serves as the specific entropy scale $S$.
In the latter case the outer boundary heat flux $q_o$ defines
$S\!=\!q_o d \alpha_o /\rho_o \kappa$ where $\kappa$ is the constant thermal
diffusivity. The magnetic field unit is
$\sqrt{\Omega\mu_0\lambda_i\rho_o}$, where $\Omega$ is the system rotation rate.

In addition to the parameters that define the background state,
Eqs.~(\ref{eq:navierstokeseq}--\ref{eq:divfieldeq}) are controlled by the Ekman
number $E$, the Rayleigh number $Ra$, the Prandtl number $Pr$ and
the magnetic Prandtl number at the inner boundary $Pm_i$:
\begin{equation}
      E=\frac{\nu}{\Omega d^2} \textrm{,}
\label{eq:ekmunber}
\end{equation}
\begin{equation}
      Ra=\frac{g_o d^3 S}{c_p \nu\kappa} \textrm{,}
\label{eq:ramunber}
\end{equation}
\begin{equation}
      Pr=\frac{\nu}{\kappa} \textrm{,}
\label{eq:prmunber}
\end{equation}
\begin{equation}
      Pm_i=\frac{\nu}{\lambda_i} \textrm{.}
\label{eq:pmmunber}
\end{equation}
To characterize the mean magnetic Prandtl number we
introduce the volume-averaged
\begin{equation}
      Pm_V = \frac{Pm_i}{V}\int_{r_i}^{r_o} 4\,\pi\,\tilde{\sigma}(r)\,r^2\, \mathrm{d}r\,
\textrm{,}
\end{equation}
where $V$ is the volume of the spherical shell.

\nad{
We will measure the dimensionless rms amplitude of the flow by the Rossby number
\begin{equation}
      Ro = E \,\left(\, \int \mathbf{u}\cdot\mathbf{u} \,dV \,\right)^{1/2}
         \textrm{ ,}
\label{eq:rossby}
\end{equation}
where $\mathbf{u}$ in this work is either only the azimuthal component of the flow velocity or the non-axisymmetric, $Ro_{zon}$ and $Ro_c$ respectively (see \tabref{TabRes}).
The analogous dimensionless rms magnetic field strength is given by the
Lorentz number,
\begin{equation}
  Lo = \left(\;\frac{E}{Pm_i} \frac{\int B^2\,dV}{\int \tilde{\rho}\,dV}\;\right)^{1/2}
  \textrm{ ,}
\end{equation}
in the form introduced by \citet{Yadav13b}.
The relative importance of the Lorentz force compared to the Coriolis force is typically quantified by the Elsasser number,
\begin{equation}
      \Lambda = \int \frac{\mathbf{B}^2}{\tilde{\rho}\tilde{\lambda}} \,dV
         \textrm{ .}
\label{eq:elsasser}
\end{equation}
}

Decisive for dynamo action is not the electrical conductivity or magnetic
diffusivity but the magnetic Reynolds number $Rm\!=\!U d/\lambda$, a combination with a typical flow velocity $U$. $Rm$ is a crude measure for the ratio of magnetic field production to Ohmic dissipation.
We introduce a radial dependent convective magnetic Reynolds number
\begin{equation}
      Rm_c(r) = \frac{U_c(r) d}{\lambda(r)}\mcom
\label{eq:rm}
\end{equation}
where $U_c(r)$ denotes the rms amplitude of non-axisymmetric flows at radius $r$.
We will also refer to the volume-averaged value
\begin{equation}
      Rm=\frac{3}{r_o^3-r_i^3} \bigintssss_{r_i}^{r_o}\! Rm_{c}(r)\,r^2\, \mathrm{d}r
      \textrm{.}
\label{eq:rm}
\end{equation}

\subsection{Numerical methods}
\label{numerics}

The anelastic equations \ref{eq:navierstokeseq} to \ref{eq:energyeq} are
solved with the MHD code MagIC\footnote{Freely available at \url{https://github.com/magic-sph/magic}}
\citep{Wicht02,Gastine12}.
MagIC has been benchmarked \citep{Jones11a} and is one of the fastest
codes of its class \citep{Matsui16}.
Poloidal/toroidal decompositions,
\begin{equation}
\begin{split}
      \tilde{\rho}\vec{u} & = \vec{\nabla}\times\left(\vec{\nabla}\times W \vec{e_r}\right)
                                    + \vec{\nabla}\times Z \vec{e_r} \mcom
      \\
      \vec{B} & = \vec{\nabla}\times\left(\vec{\nabla}\times C \vec{e_r}\right)
                        + \vec{\nabla}\times D \vec{e_r}  \mcom
\end{split}
\label{poltordec}
\end{equation}
are used for flow and magnetic field to fulfil the continuity equations.
Pseudo-spectral methods are then employed to solve
for the four unknown potential $W$, $Z$, $C$, $D$, for pressure $p$, and for
entropy $s$.  Derivatives are solved
in spectral space using a spherical harmonic decomposition in
longitude and latitude and Chebychev polynomials in radius.
Non-linear terms are calculated on a physical grid, however,
and fast Fourier transforms and Legendre transforms are used to
switch back and forth between from spectral to grid representations.

For the least demanding cases we used rather coarse grids with $72$ radial points and
a maximum spherical harmonic degree and order $\ell_{max}\!=\!133$.
For the most demanding simulations, $192$ radial grid points and $\ell_{max}\!=\!341$ were
required.
A comprehensive description of the numerical method
can be found in \cite{Glatz84} and \cite{Christensen15}.

\subsection{Quantifying Jupiter-likeness}
\label{conditions}

We mostly compare our numerical magnetic field results with VIP4 by \cite{Connerney98}
but will also discuss the newer JCF model by \cite{Ridley16}. VIP4 uses data from Pioneer and Voyager spacecrafts and from the Io auroral footprint to provide Gauss coefficients $g^m_\ell$ and $h^m_\ell$
up to degree $\ell\!=\!4$ and order $m\!=\!4$.
\cite{Ridley16} rely on all available mission data which in addition to Pioneer and Voyager also comprise
Ulysses and Galileo measurements and cover a period of $30$ years.
Their JCF fits these data with a Jupiter Constant Field model of degree and order seven.
Regularization helps to constrain smaller scale contributions.

The magnetic power spectrum at any radius $r$ above the dynamo region is
given by the square of the Gauss coefficients \citep{Lowes66,Lowes74}:
\begin{equation}
      P_{\ell}(r) = \sum_{m=0}^{\ell} E_{\ell m}^{\mbox{\scriptsize mag}}(r)
      = (\ell+1) \,\left( \frac{R_{\jupiter}}{r} \right)^{2\ell+4}
            \sum_{m=0}^{\ell}\; \left[ \left(g^m_\ell\right)^2+\left(h^m_\ell\right)^2 \right]
      \textrm{,}
      \label{eq:specl}
\end{equation}
where $E_{\ell m}^{\mbox{\scriptsize mag}}(r)$ is the magnetic energy contribution of
spherical harmonic degree $\ell$ and order $m$ and $R_{\jupiter}$ is Jupiter's surface radius.
We will compare surface spectra at $P_{\ell}(R_{\jupiter})$ but mostly rely on the rms surface field
contributions for a given degree and order:
\begin{equation}
      B_{\ell m} = \frac{ \sqrt{ E_{\ell m}^{\mbox{\scriptsize mag}}\,(R_{\jupiter}) } }
                                          {2\pi \,R_{\jupiter}^2}\textrm{.}
\end{equation}

Since our models are supposed to cover the whole gaseous envelope (or $99\%$ of the radius) we can directly compare VIP4 with the field at the top of our numerical models.
Comparing absolute field strengths would require us to rescale the non-dimensional simulations
to physical values as discussed in G14, \nad{who already demonstrated that the type of simulations considered here can indeed yield
Jupiter-like field amplitudes.
We come back to this discussion in \secref{jupiter} and first concentrate on the
field structure.}

\cite{Christensen10} define a single measure that attempts to quantify how closely a geodynamo simulation represents our knowledge of the geomagnetic field.
We follow this idea here and concentrate on comparing four key field characteristics.
Unlike for the Earth where we have an idea of field variations over many different time scales, VIP4 merely represents a snapshot.
To judge how close a given model comes to replicating VIP4 we quantify the similarity
for many snapshots with a rms misfit value \mis. Below we mostly discuss the
snapshot-average \mis\ but also show standard deviations to provide an idea of the variability.

The so-called dipolarity
\begin{equation}
      f_{dip} = \frac{B^2_{\ell=1,m=0}}
                     {B^2_{\ell\le4}}
\label{eq:dip}
\end{equation}
and dipole tilt
\begin{equation}
      \theta_{dip} = \arctan\Bigg(\frac{ B_{\,\ell=1,\,m=1} }{ B_{\,\ell=1,\,m=0} }\Bigg)
\label{eq:tilt}
\end{equation}
are often used to characterize the field geometry \citep{Duarte13}
and we will add the relative quadrupole and octopule field contributions to the mix.
However, in order to arrive at a more consistent misfit definition,
we rely on ratios of rms field contributions all the way and
use the following four measures:
\begin{equation}
      \bad = \frac{B_{\ell=1,m=0}}
                    {B_{\ell\le4}}
      \textrm{ ,}
\label{eq:quad}
\end{equation}
\begin{equation}
      \bed = \frac{B_{\ell=1,m=1}}
                    {B_{\ell\le4}}
      \textrm{ ,}
\label{eq:quad}
\end{equation}
\begin{equation}
      \bquad = \frac{B_{\ell=2}}
                      {B_{\ell\le4}}
\label{eq:quad}
\end{equation}
and
\begin{equation}
      \bocto = \frac{B_{\ell=3}}
                      {B_{\ell\le4}}
      \textrm{ .}
\label{eq:quad}
\end{equation}
The misfit is then given by
\begin{equation}
\begin{split}
      &\mis\ =
      \\
       &\sqrt{ \frac{\left(\bad\!-\!\bad_{\jupiter}\right)^2 \!+\!
 ´                    \left(\bed\!-\!\bed_{\jupiter}\right)^2 \!+\!
                      \left(\bquad\!-\!\bquad_{\jupiter}\right)^2 \!+\!
                      \left(\bocto\!-\!\bocto_{\jupiter}\right)^2 }{4}
         }\textrm{ ,}
\end{split}
\label{eq:rmserror}
\end{equation}
where the subscript $\jupiter$ refers to VIP4 values
listed in \tabref{tab:bestmodels}.
The (expected) absolute values of the four measures determine the
sensitivity of the misfit to the individual relative deviations.
For VIP4 the ratios \bquad\ and \bocto\ amount to roughly $30$\% of \bad\
and \bed\ to about $20$\% of \bad. Minimizing the misfits thus favours models that
agree with VIP4 in the relative axial dipole strength.

\section{Simulation results}
\label{results}

\subsection{Model selection}
\label{extracases}

For this study we have performed $53$ new simulations but also analyse
$13$ previously published models.
The parameters of the new cases are listed in
\tabref{TabRes} along with
diagnostic properties that have been averaged over at least $0.1$ viscous diffusion times.
Previously published models comprise G14 and polytropic simulations with $N_\rho\!=\!4$ and $N_\rho\!=\!5$ from \cite{Duarte13} and \cite{Duarte14}. Also included is a reproduction of the most realistic
model by \cite{Heimpel11}
which assumes a constant background density (Boussinesq approximation)
but uses a radial electrical conductivity profile.
More information can be found in the respective articles.

 A few of our new simulations are models where convection is driven by internal heat sources rather than bottom sources. The heating ratio $H=Q_i/Q_o$, listed in column $9$ of \tabref{TabRes},
indicates the different driving scenarios. $Q_i$ and $Q_o$ are the heat fluxes
through the inner and outer boundaries respectively. A value of  $H\!=\!0$ thus
means 100\% internal driving while $H\!=\!1$ stands for pure bottom driving.
Our models are either purely internal driven or dominantly bottom driven.
Internal heat sources seem more realistic for modelling the secular
cooling that drives convection in Jupiter \citep{Jones14}.
The term $\epsilon\,\tilde{\rho}$ in \eqnref{eq:energyeq} is the internal heat source density with $\epsilon$ the heating rate per mass. We explore
($\epsilon\!\sim\!1/\tilde{\rho}$), which results in a homogeneous internal heating,
but  also ($\nabla\epsilon\!=\!0$) where the heating is proportional
to density and thus increases with depth.
Column $10$ distinguishes between the two different thermal boundary conditions
we have explored: $S\!S$ or $F\!F$ stand for fixed
entropy or fixed flux at both boundaries.

The models cover three Ekman numbers ($E\!=\!10^{-4},3\!\times\!10^{-5},10^{-5}$) and
two Prandtl numbers ($Pr\!=\!0.1,1$).  The Rayleigh number is varied to
a certain extend, starting with low values that promise dipolar dynamos
and then increasing $Ra$ until dipolar solutions cease to exist.
To decide whether multipolar cases belong to the respective branch in the
bistability regime, we generally start each multipolar case with a
strong dipolar field.
The magnetic Prandtl number has also been varied in many cases in order to explore
whether a larger value would help to establish dipolar dynamo action.

\subsection{Onset of convection}
\label{onset}

To get a first idea of the impact of the background states and
system parameters we determined the onset of convection in the non-magnetic system.
\Tabref{Tab1} compares critical Rayleigh number and critical wave number for onset convection.
The values for the polytrope background models were calculated using a linear
solver developed by \cite{Jones09a}; the values for the
\Jupiter\ cases were determined by trial and error.
The differences between the 'polytrope' cases with $N_\rho\!=\!4$ and $N_\rho\!=\!5$
remain modest.
While the critical wave numbers found for \Jupiter\ models are similar
to those found for the polytrope models at the same Ekman and Prandtl number,
the critical Rayleigh numbers are about a factor five smaller for the polytrope models.
For a given Ekman and Prandtl, convection will thus be more
vigorous and likely also more small scale when the more realistic \Jupiter\
is considered.
Decreasing the Prandtl number from $Pr\!=\!1$ to $Pr\!=\!0.1$ leads to
a significant decrease in both critical wave number and Rayleigh number.
Internally and bottom heated cases have very similar wave numbers.
Since the critical Rayleigh numbers obey different definitions the
direct comparison is meaningless.
These results show that it is essential to adapt the Rayleigh number to the
other system parameters as well as to the background state.

\begin{table}[h!]
\centering
\caption{\small
Critical Rayleigh number $Ra_{c}$ and critical azimuthal wave numbers $m_{c}$.
The last three simulations are driven by internal rather than bottom heating and use
fixed flux ($F\!F$) rather than fixed entropy ($S\!S$) conditions.
}

\begin{tabular}{ccccccc}
\hline
$N_{\rho}$ & $Pr$ & $Ra_{cr}$ & $m_{cr}$ & $E$ & $H$ & $BC$ \\
\hline\hline
\vspace{-9pt}
 & & & & & & \\
$4.0$ & $1.0$ & $4.569\times 10^6$ & $49$ & $1\times10^{-4}$ & $1.0$ & $S\!S$ \\
$4.0$ & $1.0$ & $1.971\times 10^7$ & $80$ & $3\times10^{-5}$ & $1.0$ & $S\!S$ \\
\hline
\vspace{-9pt}
 & & & & & & \\
$5.0$ & $1.0$ & $5.372\times 10^6$ & $55$ & $1\times10^{-4}$ & $1.0$ & $S\!S$ \\
$5.0$ & $1.0$ & $2.168\times 10^7$ & $91$ & $3\times10^{-5}$ & $1.0$ & $S\!S$ \\
$5.0$ & $1.0$ & $1.155\times 10^8$ & $128$ & $1\times10^{-5}$ & $1.0$ & $S\!S$ \\
\hline
\vspace{-9pt}
 & & & & & & \\
$\jupiter$ & $0.1$ & $5.139\times 10^6$ & $21$ & $1\times10^{-4}$ & $1.0$ & $S\!S$ \\
$\jupiter$ & $0.1$ & $1.879\times 10^7$ & $38$ & $3\times10^{-5}$ & $1.0$ & $S\!S$ \\
$\jupiter$ & $1.0$ & $2.975\times 10^7$ & $55$ & $1\times10^{-4}$ & $1.0$ & $S\!S$ \\
$\jupiter$ & $1.0$ & $1.144\times 10^8$ & $96$ & $3\times10^{-5}$ & $1.0$ & $S\!S$ \\
$\jupiter$ & $1.0$ & $4.064\times 10^8$ & $156$ & $1\times10^{-5}$ & $1.0$ & $S\!S$ \\
\hline
\vspace{-9pt}
 & & & & & & \\
$\jupiter$ & $1.0$ & $2.76\times 10^7$ & $55$ & $1\times10^{-4}$ & $0.69$ & $S\!S$ \\ 
$\jupiter$ & $1.0$ & $1.07\times 10^8$ & $91$ & $3\times10^{-5}$ & $0.70$ & $S\!S$ \\ 
\hline
\vspace{-9pt}
 & & & & & & \\
$\jupiter$ & $0.1$ & $2.77\times 10^8$ & $22$ & $1\times10^{-4}$ & $0.0$ & $F\!F$ \\ 
$\jupiter$ & $1.0$ & $1.52\times 10^9$ & $55$ & $1\times10^{-4}$ & $0.0$ & $F\!F$ \\ 
$\jupiter$ & $1.0$ & $5.81\times 10^9$ & $98$ & $3\times10^{-5}$ & $0.0$ & $F\!F$ \\ 
\hline
\end{tabular}
\label{Tab1}
\end{table}

\subsection{Dynamo regimes}
\label{regimes}

\begin{figure}[h!]
      \centering
      \includegraphics[width=0.49\textwidth, trim={0.4cm 0.1cm 0.0cm 0cm}, clip]{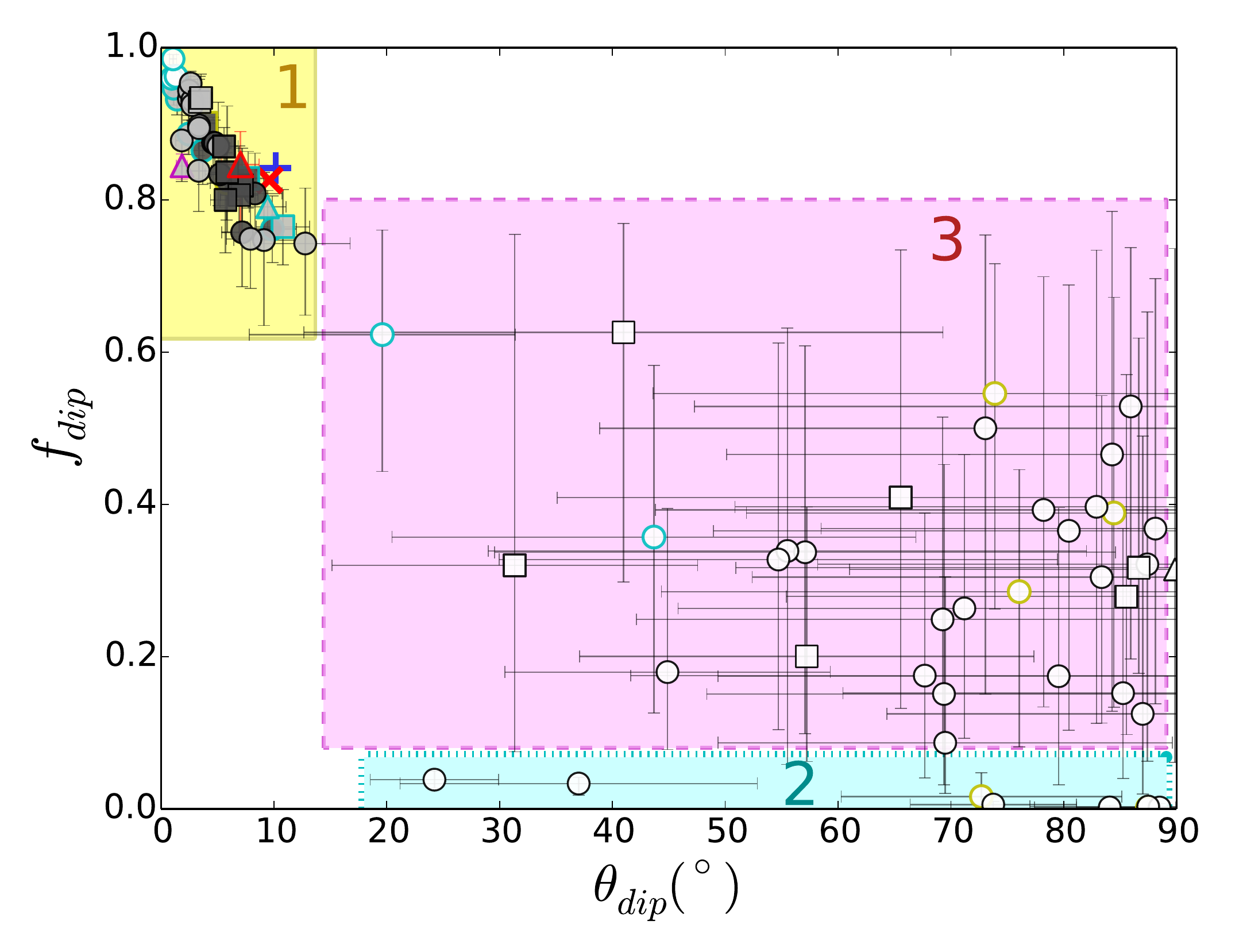}
      \caption{\small Plot of dipolarity $f_{dip}$ versus dipole tile angle $\theta_{dip}$ for the
numerical models discussed here. The colour of the symbol rim represents the
background density profile of the model (black: \Jupiter, yellow: polytrope $N_\rho\!=\!4$,
cyan: polytrope $N_\rho\!=\!5$).
The face colour of the symbols
corresponds to agreement with VIP4 quantified by \mis\: white corresponds to $\mis\!>\!0.12$, lighter grey corresponds to $0.08\!\le\!\mis\!\le\!0.12$ and darker
grey to $\mis\!\le\!0.08$ (see \eqnref{eq:rmserror}). The red '$\times$' and blue '+' symbols
show the VIP4 and JCF Jupiter field models, respectively. The red-rimmed triangle
is the G14's solution while the magenta-rimmed triangle corresponds to one simulation by
\cite{Heimpel11}. The shape of the symbols indicates the
Ekman number ($E\!=\!10^{-4}$ - circles, $E\!=\!3\times10^{-5}$ - squares,
$E\!=\!10^{-5}$ - triangles). Error bars illustrate the time
variability and show the standard deviation for values larger and smaller
than the mean independently.}
      \label{fig:diptilt1}
\end{figure}

Figure~\ref{fig:diptilt1} compares dipolarity and dipole tilt of the
numerical models with VIP4, JCF.
The solutions fall into the three distinct regimes introduced by \cite{Duarte13}.
Regime $1$, indicated by a yellow box, is characterized by
dynamos with a strong axial dipole component and generally weaker zonal flows. Regime $2$, the cyan box,
encompasses cases where the axial dipole contribution remains typically weak.
Finally, regime $3$, the magenta box, contains models where the axial dipole contribution varies strongly in time, as indicated by the larger error bars. Solutions in regimes $2$ and $3$
typically have strong zonal flows.
Several previously discovered reasons for a dynamo to end up in regime $1$ where
Jupiter-like solutions can be expected have been discussed in the introduction.
Our new simulations confirm the respective inferences and we provide a few examples in
the following. In particular the competition between zonal
flows and dipolar fields continues to play a decisive role.

The differences in the three density profiles considered here (see \figref{fig:densprofs})
have no effect on the ability of the dynamo to maintain a dipole-dominated field.
The dynamics is mainly influenced by the density gradients which differ more
significantly in the outer part of the shell.
The profiles may thus indeed yield different flows in this region but the
low conductivity layer required to guarantee dipole-dominated fields
prevent them from affecting the dynamo.

When the low conductivity layer is too thick, however, the field becomes
once more multipolar \citep{Duarte13}. Our model $12$
with a transition radius of $\chi_m\!=\!70$\% instead of $80$\% or $90$\% in the other cases
is an example for this effect.
The `early decaying' conductivity profile (yellow line in \figref{fig:vconds})
potentially also causes the same problem and most
of the respective $E\!=\!10^{-4}$ models ($9$, $35$, $45-48$)
and $E\!=\!10^{-5}$ models ($61$, $63-65$)
indeed end up being multipolar.
The remaining higher conducting region where Lorentz forces could counteract
the Reynolds stresses that drive zonal winds simply becomes too small \citep{Duarte13}.
We come back to discussing the impact of the different conductivity profiles below.

The pairs of dipolar and multipolar cases $10/11$ and $29/30$ in \tabref{TabRes}
are examples for the bistability at not too large Rayleigh numbers and $Pr\!=\!1$,
the former pair for a polytropic profile and the latter for a \Jupiter\ model.
At larger Rayleigh numbers and $Pr\!=\!1$ only the multipolar branch remains,
for example model $43$ is dipolar at $Ra/Ra_{cr}\!=\!7.4$ while only the multipolar
dynamo $49$ is found at $Ra/Ra_{cr}\!=\!8.4$.
The behaviour seems to be different at $Pr\!=\!0.1$ where we
find only multipolar solutions for $Ra/Ra_{cr}\!=\!7.8$ (model $16$)
to $Ra/Ra_{cr}\!=\!8.4$ (model $18$).
Increasing $Ra$ further to $Ra/Ra_{c}\!=\!9.8$ in model $19$, however,
establishes a dipole-dominated dynamo, a behaviour not
observed for $Pr\!=\!1$.
We have checked that dipolar solutions cannot exist at lower $Ra$ since our values in models $16-18$ represent the onset of dynamo action, but we have not checked
at which $Ra$ value multipolar solution would reappear.

\begin{figure}[h!]
\centering
      \includegraphics[height=0.25\textwidth, trim={0.0cm 0.0cm 0.0cm 0cm}, clip]{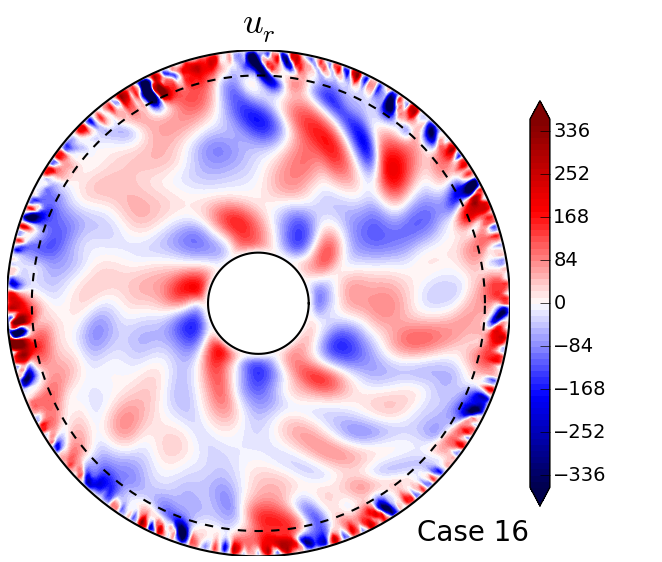}
      \includegraphics[height=0.27\textwidth, trim={0.0cm 0.0cm 0.0cm 0cm}, clip]{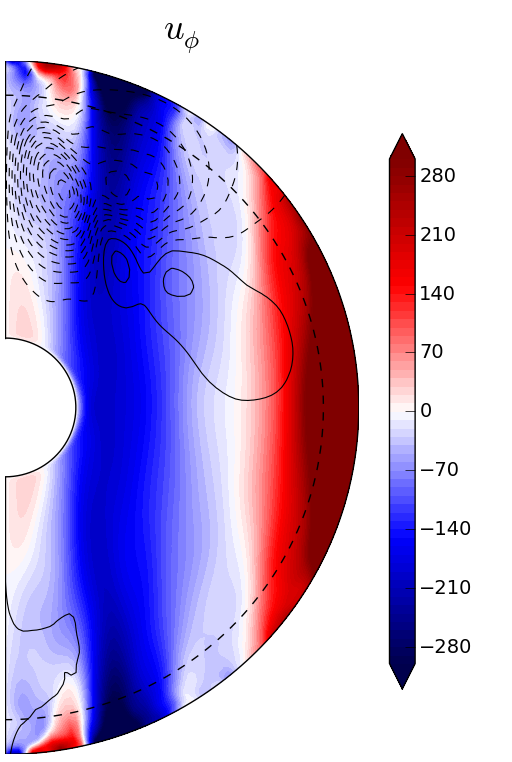}
      \\
      \includegraphics[height=0.25\textwidth, trim={0.0cm 0.0cm 0.0cm 0cm}, clip]{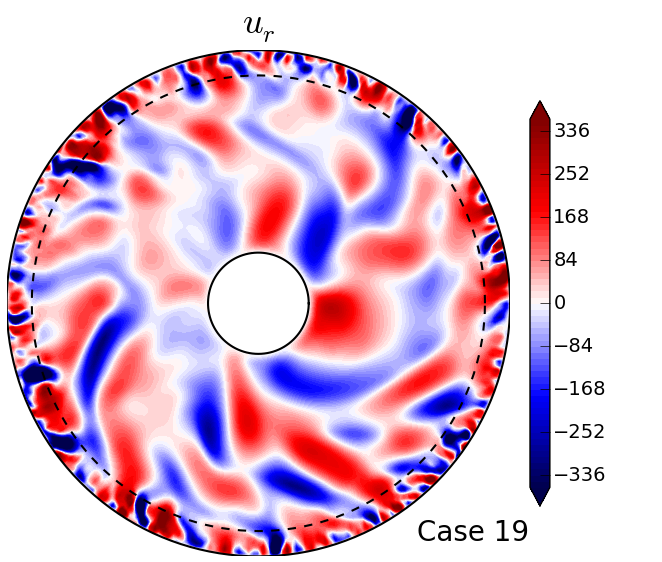}
      \includegraphics[height=0.27\textwidth, trim={0.0cm 0.0cm 0.0cm 0cm}, clip]{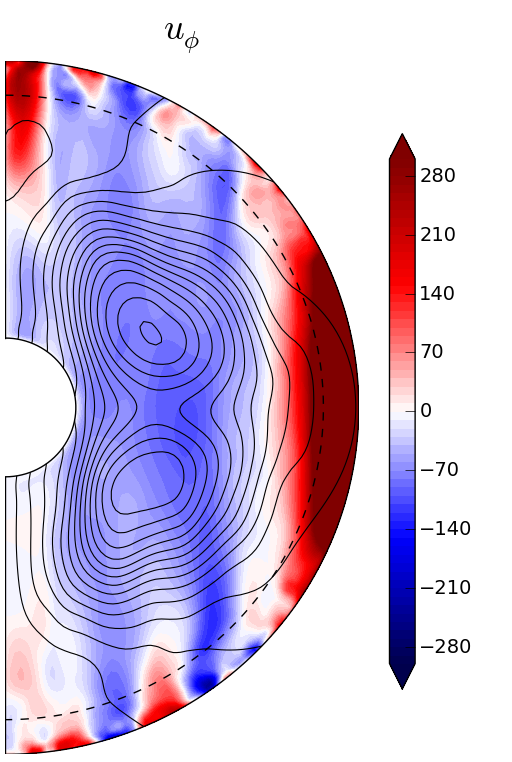}
      \\
      \includegraphics[height=0.25\textwidth, trim={0.0cm 0.0cm 0.0cm 0cm}, clip]{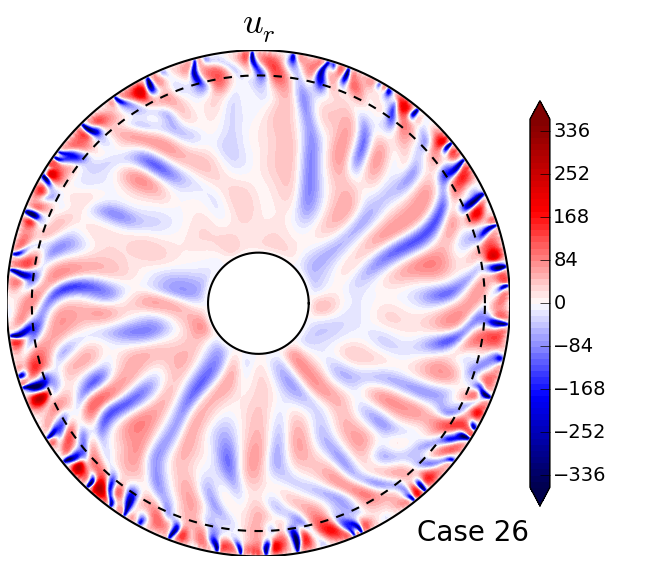}
      \includegraphics[height=0.27\textwidth, trim={0.0cm 0.0cm 0.0cm 0cm}, clip]{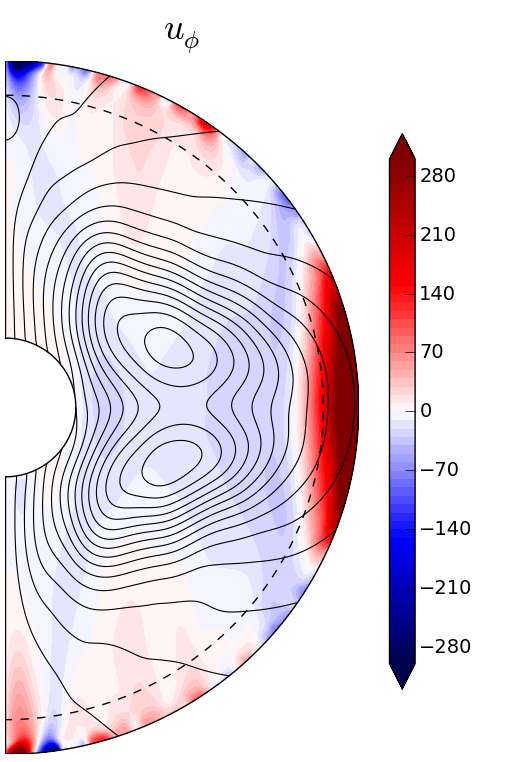}
      \\
      \includegraphics[height=0.25\textwidth, trim={0.0cm 0.0cm 0.0cm 0cm}, clip]{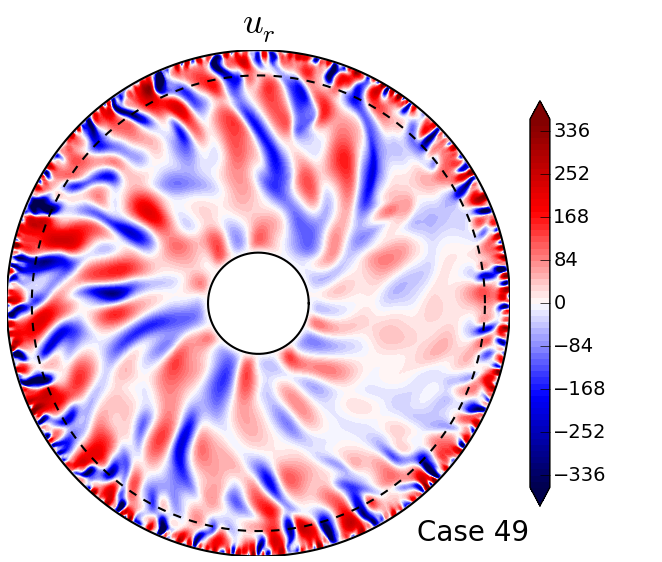}
      \includegraphics[height=0.27\textwidth, trim={0.0cm 0.0cm 0.0cm 0cm}, clip]{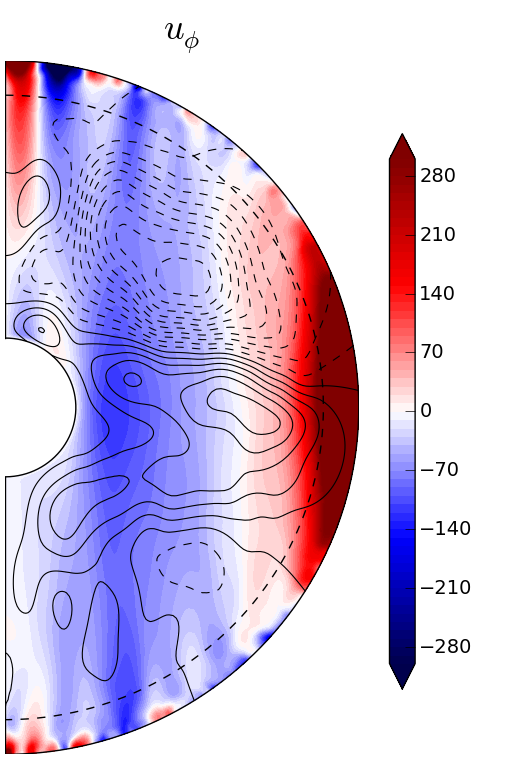}
      \caption{\small Radial flow of four models (one per row) in the equatorial plane (left column) and zonal flow with superimposed axisymmetric field lines (right column). Red (blue) indicates outward (inward) or prograde (retrograde) flows, respectively. The first two rows show cases with $Pr\!=\!0.1$ and the last two $Pr$ unity. The supercriticality is from top to bottom $7.8$, $9.7$, $5.7$, $8.4$. The dashed lines correspond to the $r_m$ parameter in the variable conductivity profile definition (see \eqnref{eq:varcondeq}).}
\label{fig:flow}
\end{figure}

\Figref{fig:flow} compares dipolar and multipolar simulations which differ
only in Prandtl and Rayleigh number. The $Pr\!=\!1$ cases
$26$ and $49$ ran at $Ra/Ra_{cr}\!=\!5.7$ and $Ra/Ra_{cr}\!=\!8.4$
while the  $Pr\!=\!0.1$ models $16$ and $19$ have
$Ra/Ra_{cr}\!=\!7.8$ and $Ra/Ra_{cr}\!=\!9.7$, respectively.
The non-axisymmetric flows,
illustrated with radial flow contours in the left column,
are generally larger scale for $Pr\!=\!0.1$ than for $Pr\!=\!1$
as already predicted by the critical wave numbers (see \tabref{Tab1}).
Note that the scale difference is more pronounced in the deeper region
than in the weakly conducting outer shell delineated by a dashed line.

The smaller $Ra$ flows ($16$ and $26$) show larger differences with
a much stronger increase of radial velocities for $Pr\!=\!1$.
Also, the strong dipolar field at $Pr\!=\!1$ strongly suppresses the
inner retrograde zonal jet (right column of \figref{fig:flow}.
The faster non-axisymmetric deeper flows in the $Pr\!=\!0.1$ solution
actually drive a surprisingly vigorous inner jet, much faster
than even in a non-magnetic $Pr\!=\!1$ case \citep{Gastine12}.

\Figref{fig:flow} shows that in addition to the strong geostrophic equatorial jet there are also shallower mid- to high-latitude zonal winds that remain confined to the weakly conducting region. These winds are somewhat more pronounced in the low Prandtl number simulations $16$ and $19$ but they are never much faster than the non-axisymmetric flow contributions. Moreover, while the equatorial jet is very persistent over the whole duration the mid- to high-latitude winds change on time scales only slightly longer than the
typical time scale of the non-axisymmetric flows (of comparable length scale). Both type of zonal flows features therefore seem to obey different dynamics but further in-depth analysis is required clarify whether this may change at more realistic parameters.

For both Prandtl numbers, the radial gradient in non-axisymmetric
flow velocities decreases with Rayleigh number.
Both larger $Ra$ flows ($19$ and $49$) then seem rather similar.
Except for the scale difference there is no apparent reason why only one of the
two supports a dipole-dominated dynamo.
We can only conclude that the competition between the dipolar field
and Reynolds stress driven zonal winds is less severe for $Pr\!=\!0.1$
once the deeper flows become stronger at larger $Ra$.

Our new $Pr\!=\!1$ simulations confirm the trade-off between
the Rayleigh number and the magnetic Prandtl described by
various authors \citep{Duarte13,Schrinner14,Jones14,Raynaud15}:
an increase in $Pm$ can prevent a dynamo from becoming multipolar,
Cases $32$ to $34$ illustrate this effect:
case $32$ is multipolar at $Pm_i\!=\!1$ while cases $33$ and $34$ are dipolar at
$Pm_i\!=\!2$ and $Pm_i\!=\!3$, respectively.
Cases $42$ and $43$ are another example where increasing
$Pm_i$ from $1$ to $2$ proved successful.
We expect that increasing $Pm$ should also extend the regime
of dipolar dynamo action at $Pr\!=\!0.1$ as well at least to a certain degree.

All of our $9$ purely internally heated models are multipolar which confirms the
conjecture by \cite{Jones14} that
this driving mode favours complex magnetic field configurations.
Even the choice of a small Prandtl number $Pr\!=\!0.1$ and a large
magnetic Reynolds number of up to $Pm_i\!=\!8$ did not help to
recover dipole-dominated dynamo action. Note that \cite{Jones14}
normalizes $Pm$ by the mid-depth value while we use the inner boundary value.

Whether or not a lower Ekman number helps to stabilize
dipole-dominated dynamo action is hard to conclude from our data set.
Though we have ran cases at three different Ekman numbers  they also often
differ in the other system parameters, which makes it difficult to
isolate the Ekman number effect.
An indication may be that we now find two cases ($60$ and $62$) where an
'early decaying' profile yields dipole-dominated dynamo action.
This is far from being conclusive and additional simulations are
required to clarify this point.

\subsection{Dipolar models}
\label{dipoles}

\begin{figure}[h!]
\centering
      \includegraphics[width=0.486\textwidth, trim={0.4cm 0.6cm 0.0cm 0.4cm}, clip]{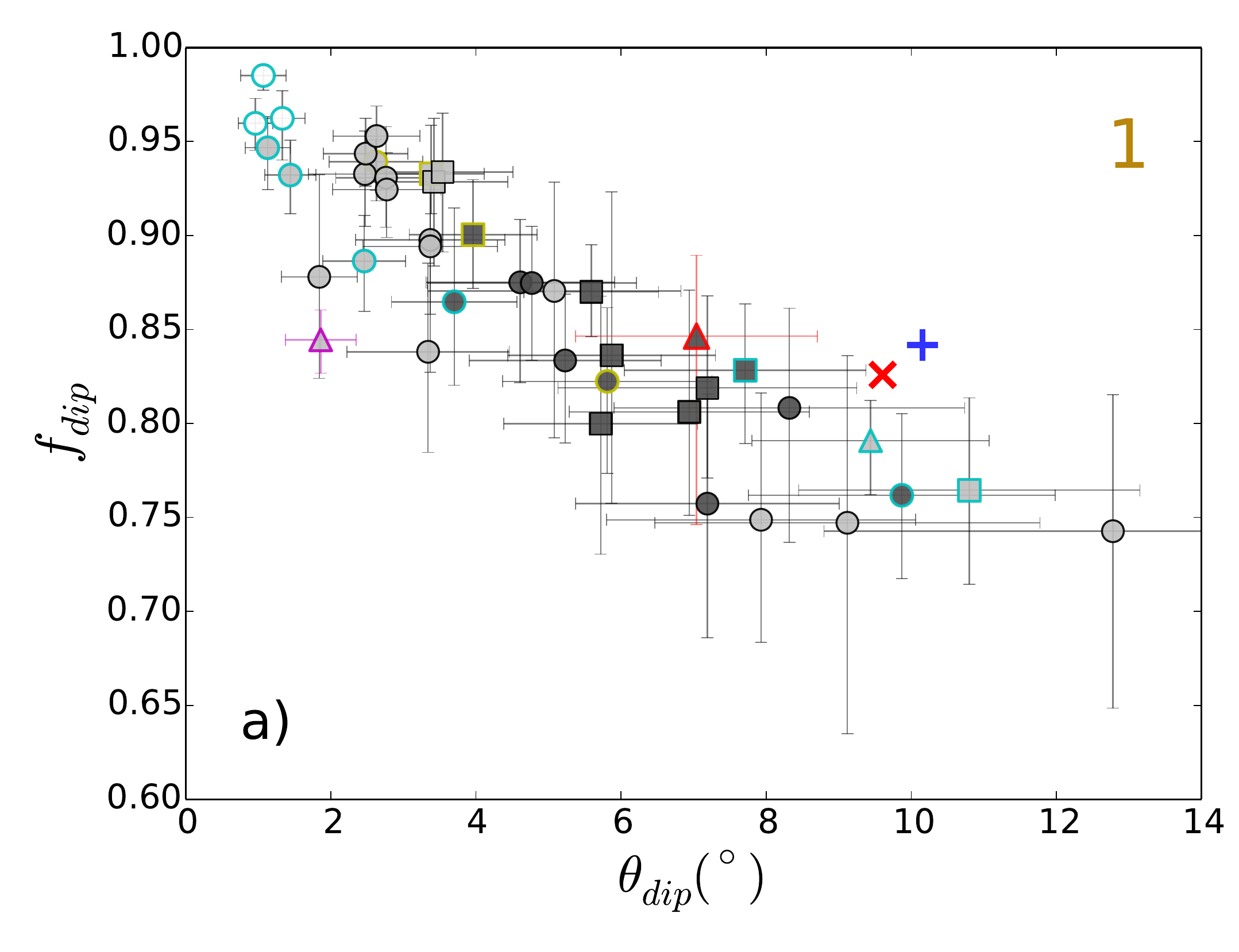}
      \includegraphics[width=0.486\textwidth, trim={0.4cm 0.6cm 0.0cm 0.4cm}, clip]{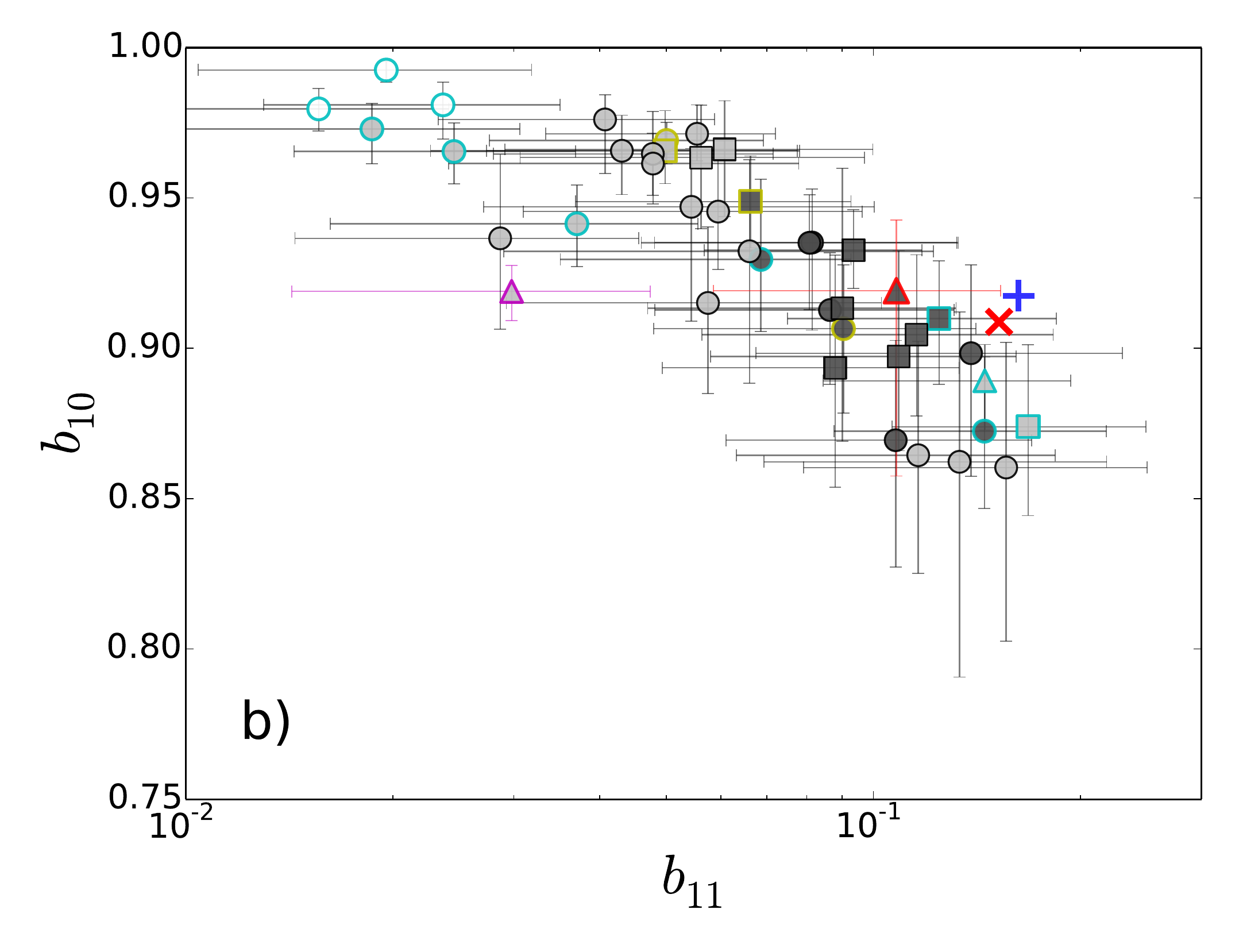}
      \includegraphics[width=0.486\textwidth, trim={0.4cm 0.6cm 0.0cm 0.4cm}, clip]{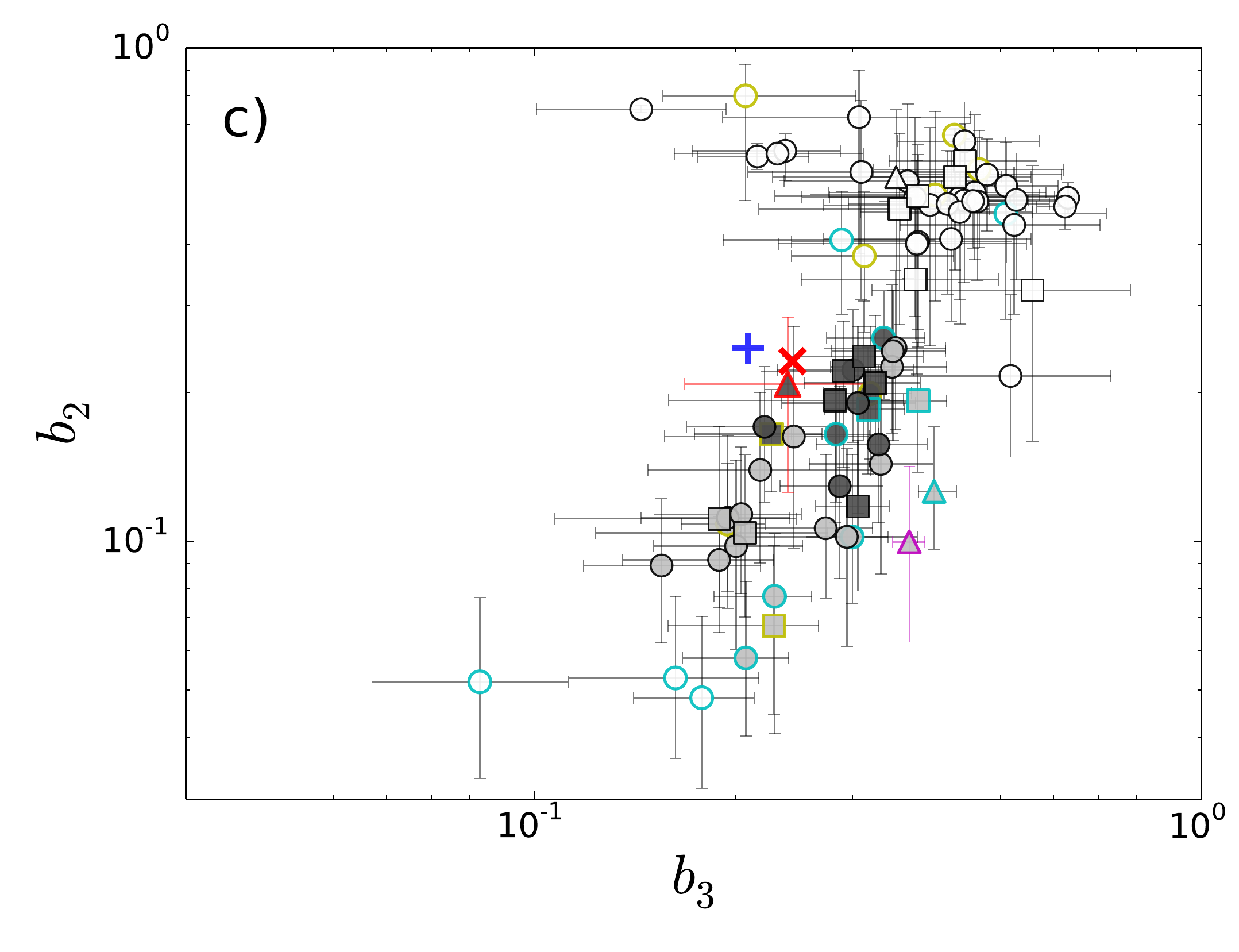}
\caption{\small Top panel: axial dipole contribution $f_{dip}$ (related to \bad\ through \eqnref{eq:dip}) and dipole tilt angle $\theta_{dip}$ (related to \bed\ through \eqnref{eq:tilt}) for the dipole-dominated solutions in regime $1$ of \figref{fig:diptilt1}. The red `$\times$' symbol corresponds to VIP4 and the blue `$+$' to JCF. Symbols represent time averages while the error bars show standard deviations (see \figref{fig:diptilt1} for more explanation). Middle and bottom panels: axial dipole contribution \bad\ and equatorial dipole contribution \bed\ in the middle panel, and relative quadrupole \bquad\ and octupole contributions \bocto\ in he bottom panel.}
\label{fig:diptilt}
\end{figure}

We proceed with discussing how well the dipolar solutions replicate VIP4 or JCF.
\Figref{fig:diptilt}a shows dipolarity and tilt for all models in regime $1$
and demonstrates that both are correlated in the sense that strongly dipolar models also tend
to have lower tilts. At least to a large degree this correlation
simply reflects the definition of the shown field characteristics.
Both mean relative dipole contributions are already very similar to
the VIP4 and JCF counterparts for many of our dipolar models.
Dark symbols in \figref{fig:diptilt} indicate particularly Jupiter-like models
with $\mis$ values below $0.08$. The misfit values of all out models
are listed in the last column \tabref{TabRes}.
The 'error' ellipses spanned by the standard deviations for some of our best models
actually include VIP4 and JCF
which means that the respective Jupiter values are closely recovered at times.
A slightly larger equatorial dipole would further reduce the average misfit.
A tilt of $\theta\!\sim\!14^\circ$ marks the boundary between dipolar and multipolar dynamos
which also seems to be true for geodynamo simulations \citep{Wicht10}.
Jupiter's dynamo seems to operate within that boundary as well.

Figures~\ref{fig:diptilt}b,c show the relative quadrupole
and octupole contributions, respectively, for all considered models. Multipolar cases reach values beyond $0.3$
and some come even close to equipartition at $b_2\!\approx\!b_3\!\approx\!0.4$,
indicating comparable magnetic energy in each spherical harmonic degree.
VIP4 or JCF values are once more inside the 'error ellipses' for a number of our best models but
a smaller relative octupole contribution would further improve the overall agreement.
Note that the triangle for the G14 model lies particularly close to VIP4.

\begin{figure}[h!]
\centering
      \includegraphics[width=0.485\textwidth]{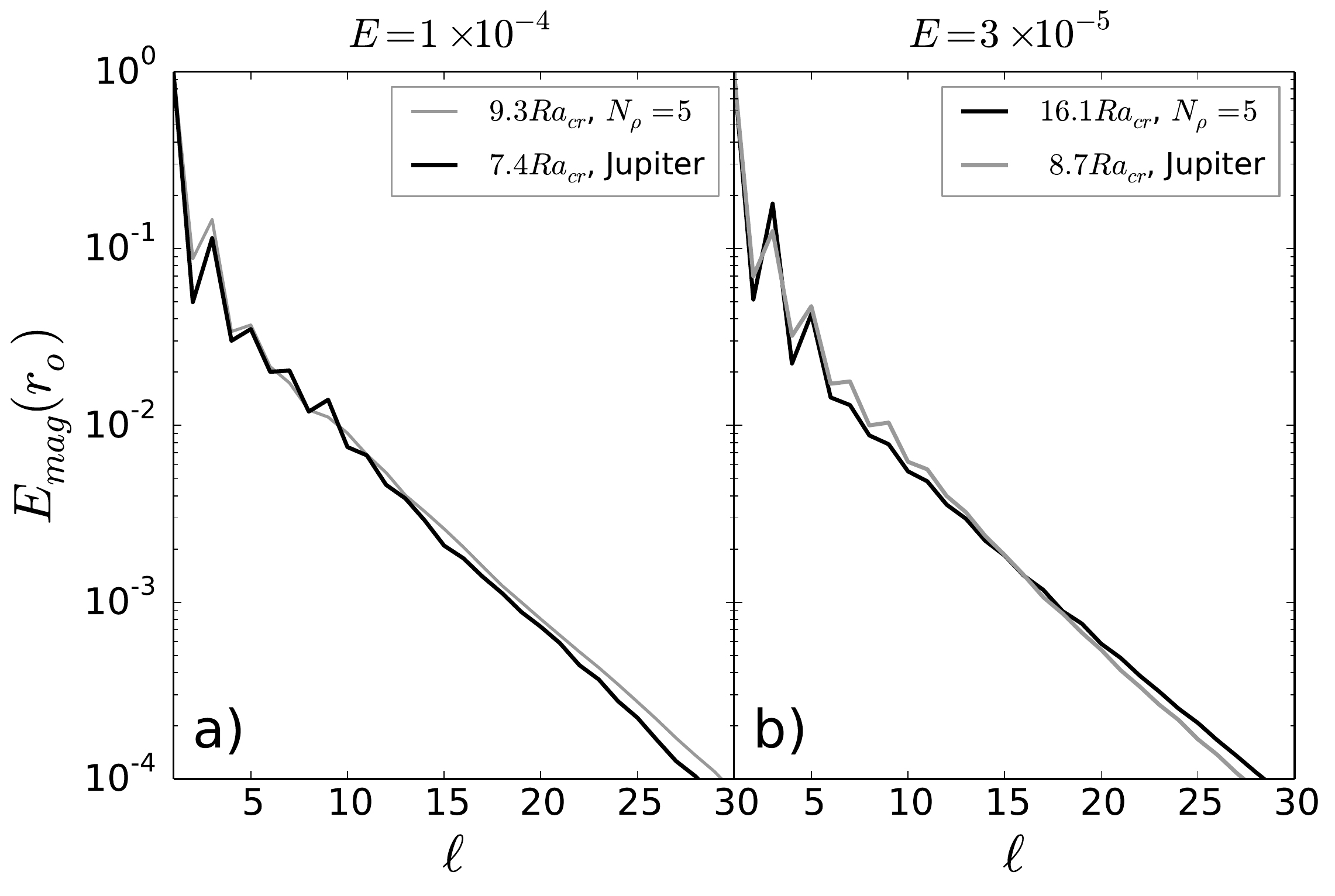}
\caption{\small Magnetic spectra as a function of the spherical harmonic degree $\ell$ for two different density profiles and the G14 electrical conductivity model. Grey lines show simulations using the G14 density model while black lines indicate $N_\rho\!=\!5$ polytrope models. The left panel shows cases $8$ and $43$ with $E\!=\!10^{-4}$ and the right panel
cases $14$ and $59$ with $E\!=\!3\times10^{-5}$.}
\label{fig:denspecs}
\end{figure}

Figures~\ref{fig:diptilt}b,c suggest that neither
the Ekman numbers (symbol type) nor the background density models (rim colour)
have a direct impact on how closely a dipole-dominated simulation replicates VIP4.
\Figref{fig:denspecs} demonstrates the similarity of the magnetic spectra
for two dipolar solutions with a polytropic $N_\rho\!=\!5$
and the \Jupiter\ density at two different Ekman numbers.
As discussed above, the electrical conductivity profiles chosen in this work likely
reduce the potential impact of the density model.

The simulations reveal two other important effects that decisively influence the relative
spectral contributions in our dipolar dynamo solutions.
Several of our magnetic fields are actually too dipolar. The axial dipole contribution is so dominant that these models end up as a cluster in the upper left part of \figref{fig:diptilt}a and the lower left part of the bottom panel.
Comparing, for example, models $4$ and $8$ in Tab.~\ref{TabRes} illustrates
this effect: while in model $4$ the axial dipole is far too dominant,
increasing $Ra$ by $25$\% leads to a very VIP4-like model $8$.
 In many cases we could identify a
too small Rayleigh number as the reason.
Another consequence of a smallish Rayleigh number is often a too simplistic
time dependence which, however, is not considered here \citep{Wicht10}.

\begin{figure}[h!]
\centering
      \includegraphics[width=0.492\textwidth]{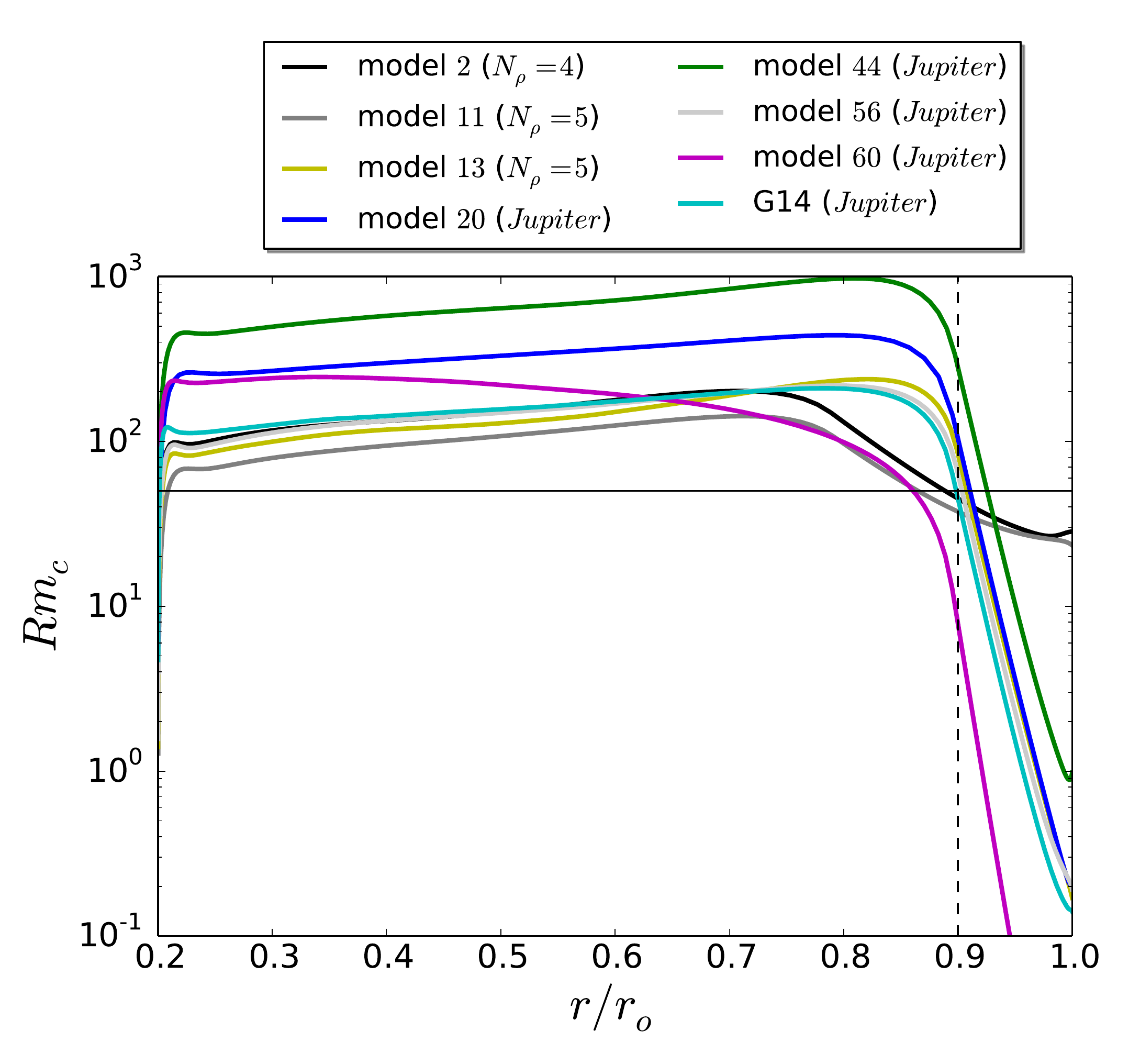}
\caption{\small Convective magnetic Reynolds number profiles (Eq.~\ref{eq:rm}) for seven new simulations are for G14. See \tabref{tab:bestmodels} and \tabref{TabRes} for more information on the model parameters and properties. The horizontal black line marks $Rm_c\!=\!50$ and the vertical dashed line $r/r_o\!=\!0.9$.}
\label{fig:rmsr}
\end{figure}

The second important factor that influences how close our simulations come to
replicating Jupiter's magnetic field is the electrical conductivity model.
The combination of electrical conductivity and flow amplitude determines
the magnetic Reynolds number profile $Rm_c(r)$ and thus the depth of the dynamo process.
Since the flow profiles typically increase mildly with radius, it is mostly
the decrease in $Rm_c$ that delimits the dynamo region.
\Figref{fig:rmsr} illustrates $Rm_c(r)$
for six of our best models ($2,11,13,20,44,56$) and
one less successful example ($60$).
We have added a horizontal line indicating the critical value $Rm\!=\!50$ where
dynamo action becomes possible in Boussinesq simulations \citep{Christensen06}.
The normalized radius beyond which $Rm_c$ remains below $50$ will be referred to
as $\RD$ in the following: $Rm_c(r)\!<\!50\;\mbox{for}\;r/r_o\!\ge\!\RD$.

Most simulations that come closer to replicating VIP4 use the G14 electrical conductivity profile
(cyan line in \figref{fig:vconds}) like models $13$, $20$, $44$ and $56$ depicted in \figref{fig:rmsr}.
The profiles for $13$, $56$, and $G14$ are very similar and all yield $\RD\!=\!r/r_o\!=\!0.9$.
The high magnetic Prandtl number of $Pm_i\!=\!10$ in model $44$ leads to
the largest magnetic Reynolds numbers of up to $Rm_c\!=\!1000$ in our simulations.
The respective green profile in \figref{fig:rmsr} illustrates that
$\RD$ is pushed out to a particularly large value of $0.92$.
The blue line shows model $20$, which ranks second in
terms of $Rm$ amplitudes and has a slightly smaller $\RD$ value of $0.91$.

The magenta line in \figref{fig:rmsr} shows the profile for model $62$ which uses
the early decaying conductivity model (cyan line in \figref{fig:vconds}).
Dynamo action is then more concentrated at depth with a smallish $\RD\!=\!0.86$.
Most of the simulations using this profile yield multipolar dynamos and
those which remain dipolar (models $9,60,62$) have a too strong axial dipole
and end up as white or light grey dots in the upper left corner of \figref{fig:diptilt}a and the lower left part of \figref{fig:diptilt}c.
The reason is likely purely geometric: assuming that the field for radii $r\!>\!\RD$ is
a potential field the individual spherical harmonic surface field contributions
are by a factor $(\RD/r_o)^{\ell+2}$ smaller than at $\RD$.
The dominance of the dipole contribution thus increases with decreasing $\RD$.

Models using electrical conductivity profiles with a deeper
transition radius, for example $\chi_m\!=\!80\%$ instead of $\chi_m\!=\!90\%$,
face the same principal problem (models $1,2,3,5-7,11$ and $30,31,40$)
but the decay rate naturally also plays a role.
Models $2$ and $11$ use $\chi_m\!=\!80\%$ but combined with a
mild conductivity drop of only two orders of magnitude (black line in in \figref{fig:vconds}).
Black and grey lines in \figref{fig:rmsr} illustrate the respective $Rm_c$ profiles.
The larger flow velocities in model $2$ push the dynamo region out to $\RD\!=\!0.89$
while $\RD\!=\!0.86$ in model $11$. Both actually have very favourable misfit values
of $\mis\!=\!0.065$ and $\mis\!=\!0.067$, respectively, despite the fact that the electrical conductivity profiles are not very Jupiter-like. We will demonstrate below
that higher harmonic field contributions should allow to dismiss these models.

\begin{figure}[h!]
      \centering
      \includegraphics[width=0.493\textwidth, trim={0.4cm 0.1cm 0.0cm 0cm}, clip]{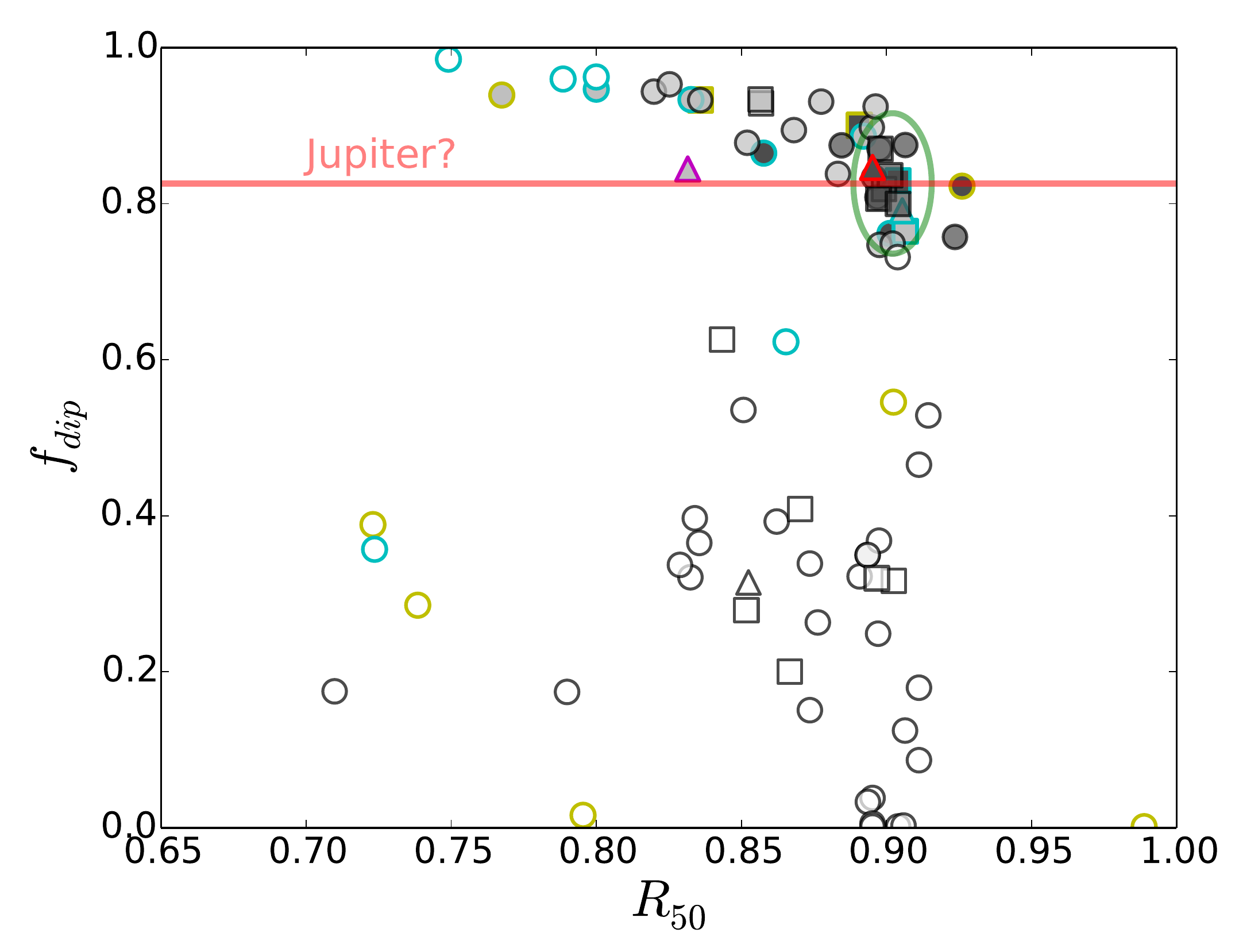}
      \caption{\small{Relative axial dipole contribution plotted against
$\RD$, the relative radius where the magnetic Reynolds number drops below $50$.
See \figref{fig:diptilt} for an explanation of the different symbols.}
}
\label{fig:radRm50}
\end{figure}

\Figref{fig:radRm50} shows the dependence of the relative axial dipole contribution
\bad\ on \RD\ for all analysed dynamo models.
Model $56$ with $\RD\!=\!0.90$ and $\bad\!=\!0.12$ has the smallest misfit value of
$\mis\!=\!0.061$. We generally regard the dark-coloured cases with $\mis\!\le\!0.08$ as our 'best models'.
They cluster around $\RD\!\approx\!0.9$ in the region
marked by an ellipse in \figref{fig:radRm50}.
However, since the two models with the largest values of
$\RD\!=\!0.93$ (model $1$) and $\RD\!=\!0.92$ (model $44$) are among our most Jupiter-like cases
the upper bound remains unclear.
Exploring models with yet larger \RD\ radii should clarify this point in the future.

Given the large internal magnetic Reynolds numbers in Jupiter, the
weak conductivity decrease predicted for the metallic layer
practically plays no role. However, our simulations for the
slowly decaying conductivity profile that attempts to replicate this
feature has shown that it strongly influences
the numerical dynamo models.
For the limited flow velocities in our dipolar models, the
conductivity profile used by G14 and our best models seems to offer
a good compromise.

\begin{table*}[h!]
{\footnotesize
\centering
\begin{tabular}{ccccccccccccccccc}
\hline
\vspace{-8.5pt}
 & & & & & & & & & \\
Model & $N_{\rho}$ & $E$ & $\dfrac{Ra}{Ra_{cr}}$ & $Pm_i$ & $Pm_V$ & $Pr$ & $\begin{matrix}r_{50}\\(\%)\end{matrix}$ & $\begin{matrix}f_{dip}\\ \ell\le4\end{matrix}$ & $\begin{matrix}\theta_{dip}\\(^\circ)\end{matrix}$ & $\dfrac{ B_{\,\ell=1,\,m=1} }{ B_{\,\ell\le4} }$ & $\dfrac{ B_{\,\ell=2} }{ B_{\,\ell\le4} }$ & $\dfrac{ B_{\,\ell=3} }{ B_{\,\ell\le4} }$ & $\begin{matrix}Lo(r_o)_{\,\ell\le4}\\10^{-2}\end{matrix}$ & $\begin{matrix}Lo\\10^{-2}\end{matrix}$ & $\begin{matrix}Ro_c\\10^{-2}\end{matrix}$ & $\mis$ \\
\vspace{-8pt}
 & & & & & & & & & \\
\hline\hline
\vspace{-10pt}
 & & & & & & & & & \\
 \rowcolor{lightgray} VIP4 & $\sim\!8$ & $\sim\!5\!\times\!10^{-19}$ & $-$ & $-$ & $-$ & $-$ & $-$ & $0.83$ & $9.6$ & $0.152$ & $0.232$ & $0.244$ & $-$ & $-$ & $-$ & $-$ \\
 RID & $\sim\!8$ & $\sim\!5\!\times\!10^{-19}$ & $-$ & $-$ & $-$ & $-$ & $-$ & $0.85$ & $10.2$ & $0.163$ & $0.247$ & $0.210$ & $-$ & $-$ & $-$ & $-$ \\ 
\hline
 $1$ & $4$ & $10^{-4}$ & $5.5$ & $4.0$ & $2.06$ & $1.0$ & $92.6$ & $0.82$ & $5.8$ & $0.091$ & $0.199$ & $0.319$ & $0.134$ & $0.93$ & $1.57$ & $0.067$ \\ 
 \rowcolor{lightgray} $2$ & $4$ & $3\!\times\!10^{-5}$ & $5.6$ & $2.0$ & $1.03$ & $1.0$ & $89.1$ & $0.90$ & $4.0$ & $0.066$ & $0.165$ & $0.227$ & $0.101$ & $0.53$ & $0.75$ & $0.065$ \\ 
 $8$ & $5$ & $10^{-4}$ & $9.3$ & $2.0$ & $1.03$ & $1.0$ & $90.1$ & $0.76$ & $9.9$ & $0.145$ & $0.258$ & $0.334$ & $0.096$ & $0.89$ & $2.15$ & $0.070$ \\ 
 \rowcolor{lightgray} $11$ & $5$ & $10^{-4}$ & $9.3$ & $2.0$ & $1.03$ & $1.0$ & $85.8$ & $0.87$ & $3.7$ & $0.069$ & $0.165$ & $0.283$ & $0.135$ & $0.96$ & $2.08$ & $0.067$ \\ 
 \rowcolor{lightgray} $13$ & $5$ & $3\!\times\!10^{-5}$ & $9.2$ & $1.5$ & $0.96$ & $1.0$ & $90.4$ & $0.83$ & $7.7$ & $0.125$ & $0.185$ & $0.316$ & $0.048$ & $0.44$ & $1.02$ & $0.061$ \\ 
 \rowcolor{lightgray} $20$ & $\jupiter$ & $10^{-4}$ & $11.7$ & $2.0$ & $1.27$ & $0.1$ & $90.7$ & $0.88$ & $4.6$ & $0.081$ & $0.171$ & $0.221$ & $0.124$ & $2.25$ & $3.95$ & $0.064$ \\ 
 $33$ & $\jupiter$ & $10^{-4}$ & $6.1$ & $2.0$ & $1.27$ & $1.0$ & $89.6$ & $0.83$ & $5.2$ & $0.087$ & $0.157$ & $0.328$ & $0.110$ & $0.89$ & $1.55$ & $0.077$ \\ 
 $39$ & $\jupiter$ & $10^{-4}$ & $6.1$ & $2.0$ & $1.27$ & $1.0$ & $88.5$ & $0.88$ & $4.8$ & $0.081$ & $0.129$ & $0.287$ & $0.151$ & $0.95$ & $1.55$ & $0.078$ \\ 
 \rowcolor{lightgray} $43$ & $\jupiter$ & $10^{-4}$ & $7.4$ & $2.0$ & $1.27$ & $1.0$ & $89.7$ & $0.81$ & $8.3$ & $0.139$ & $0.191$ & $0.306$ & $0.084$ & $0.83$ & $1.93$ & $0.074$ \\ 
 $44$ & $\jupiter$ & $10^{-4}$ & $7.4$ & $10.0$ & $6.39$ & $1.0$ & $92.4$ & $0.76$ & $7.2$ & $0.108$ & $0.222$ & $0.301$ & $0.053$ & $1.06$ & $2.13$ & $0.067$ \\ 
 $55$ & $\jupiter$ & $3\!\times\!10^{-5}$ & $7.4$ & $1.2$ & $0.77$ & $1.0$ & $89.8$ & $0.87$ & $5.6$ & $0.094$ & $0.118$ & $0.306$ & $0.068$ & $0.47$ & $0.87$ & $0.076$ \\ 
 \rowcolor{lightgray} $56$ & $\jupiter$ & $3\!\times\!10^{-5}$ & $7.4$ & $1.5$ & $0.96$ & $1.0$ & $89.9$ & $0.82$ & $7.2$ & $0.116$ & $0.221$ & $0.291$ & $0.057$ & $0.48$ & $0.87$ & $0.061$ \\ 
 $57$ & $\jupiter$ & $3\!\times\!10^{-5}$ & $7.9$ & $1.5$ & $0.96$ & $1.0$ & $90.1$ & $0.84$ & $5.9$ & $0.090$ & $0.193$ & $0.282$ & $0.061$ & $0.50$ & $0.91$ & $0.078$ \\ 
 $58$ & $\jupiter$ & $3\!\times\!10^{-5}$ & $8.7$ & $1.2$ & $0.77$ & $1.0$ & $89.7$ & $0.81$ & $6.9$ & $0.109$ & $0.209$ & $0.324$ & $0.062$ & $0.52$ & $1.05$ & $0.070$ \\ 
 $59$ & $\jupiter$ & $3\!\times\!10^{-5}$ & $8.7$ & $1.5$ & $0.96$ & $1.0$ & $90.4$ & $0.80$ & $5.7$ & $0.088$ & $0.237$ & $0.312$ & $0.054$ & $0.52$ & $1.03$ & $0.072$ \\ 
\hline
 \rowcolor{lightgray} G14 & $\jupiter$ & $10^{-5}$ & $12.3$ & $0.6$ & $0.37$ & $1.0$ & $89.5$ & $0.85$ & $7.0$ & $0.108$ & $0.208$ & $0.240$ & $0.038$ & $0.27$ & $0.69$ & $0.063$ \\
 CAJ & $\jupiter$ & $2.5\!\times\!10^{-5}$ & $-$ & $6.2$ & $1.06$ & $0.1$ & $-$ & $0.91$ & $7.4$ & $0.123$ & $0.206$ & $0.155$ & $-$ & $-$ & $-$ & $0.053$ \\ 
\hline
\end{tabular}
\caption{\small{Numerical models that best match the VIP4
observational model\citep{Connerney98}, according to the results of Eq.~\ref{eq:rmserror}. We also show the more recent model from \cite{Ridley12}. For
comparison purposes, the G14 model has also been included as well as the model and single snapshot in Fig.5c of \cite{Jones14}.
}}
\label{tab:bestmodels}
}
\end{table*}

The parameters and field characteristics of our $15$ best cases
are listed in \Tabref{tab:bestmodels} along with
the values for G14 and the observational models VIP4 and JCF
(\tabref{TabRes} provides an overview for all models).
The relative axial dipole contributions (column $8$) of the selected numerical models
closely resemble Jovian values and range between $92$\% and $108$\% of the VIP4 data.
The other characteristics vary more strongly,
the relative equatorial dipole (column $11$)
between $43$\% and $95$\% , the relative quadrupole (column $12$)
between $51$\% and $111$\%  and the relative octupole between $91$\%
and $137$\% of the respective VIP4 values.
As already discussed above, relative equatorial dipole contributions or quadrupole
contributions are on average somewhat smaller than the Jovian values while
the octupole is larger, but the overall agreement is indeed very decent.

The fact that the best models have different Ekman numbers, Prandtl numbers,
and background models
once more illustrates that Jupiter-like solutions can be found over a broader range of
parameters, at least when Rayleigh number and magnetic Prandtl number are
adjusted accordingly.

We should point out that the axial dipole and octupole contributions have opposite signs both in our models and in VIP4. This means a stronger field concentration nearer the equator (latitudes between $\pm 45^\circ$) and weaker at higher latitudes toward the polar regions. We will come back to this discussion in the following.

\subsection{Beyond the $\mis$ measure}
\label{jmodels}

\begin{figure}[h!]
\centering
      \includegraphics[width=0.5\textwidth, trim={0.4cm 0.6cm 0.0cm 0.4cm}, clip]{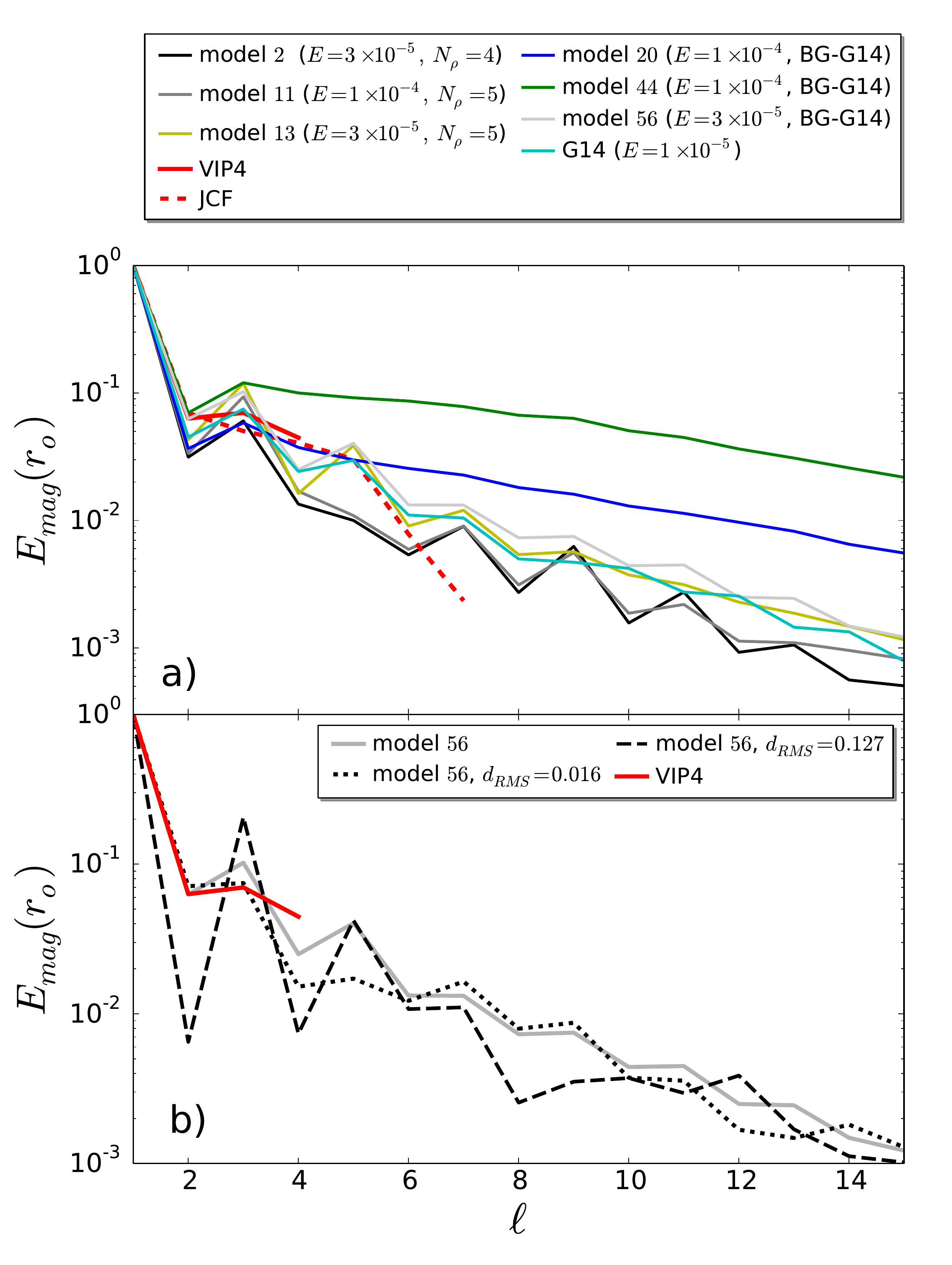}
\caption{
\small (a) Time-averaged surface field spectra for six of the best numerical models, and for G14 and VIP4. Line colours refer to the same models as in \figref{fig:rmsr}. More information on the models can be found in \tabref{tab:bestmodels}.
(b) Comparison of spectra for the numerical model $56$ with VIP4 (red line). Model $56$ has the smallest mean $\mis$ value of all considered dynamos. While the grey solid line shows the mean spectrum for model $56$, dotted and dashed lines depict spectra for snapshots with particularly low (dotted) or large (dashed) $\mis$ values assumed during the model evolution.
}
\label{fig:bestspecs}
\end{figure}

So far we have discussed the Jupiter-likeness of our numerical simulations in terms of
our misfit measure $\mis$ where only the first three spherical harmonics enter in
an averaged sense.
We now take a brief look at the higher harmonic contributions, at the time dependence, and
at surface field maps.

\Figref{fig:bestspecs}a shows the normalized magnetic power spectra up to degree $\ell\!=\!14$ for six of our best models, along with VIP4, JCF and G14.
At least the $\ell\!=\!6$ and $7$ contributions in JCF are likely strongly controlled by the
regularization and are not considered further.
Though all the numerical models have similarly small
$\mis$ values, four of the numerical spectra show distinct differences.
Models $2$ and $11$ use the slowly decaying conductivity model
(black line in in \figref{fig:vconds}) which results in
a mildly decaying magnetic Reynolds number profile
(black and grey lines in \figref{fig:rmsr}).
The respective spectra show particularly low $\ell\!=\!4$, $5$ and $6$ contributions
which seem incompatible with the Jovian field.

The remaining numerical models use the same conductivity profile of G14.
Models $20$ and $44$ reach particularly large
magnetic Reynolds numbers which results in
generally stronger higher harmonic contributions and an
overall smoother spectrum.
Already the $\ell\!=\!4$ contribution of model $44$ seems too high
but model $20$ values are well acceptable.

\begin{figure*}[h!]
\centering
      \includegraphics[height=3.6cm, trim={0cm 0 0 1.3cm}, clip]{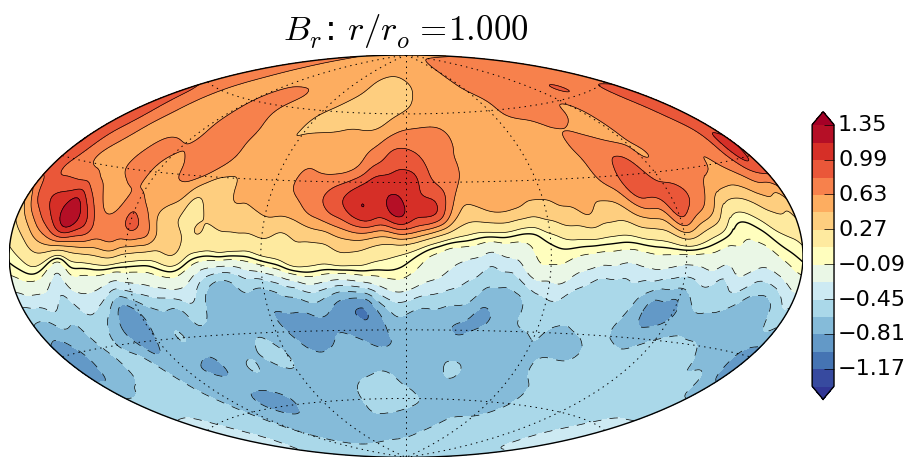} 
      \includegraphics[height=3.6cm, trim={0cm 0 0 1.3cm}, clip]{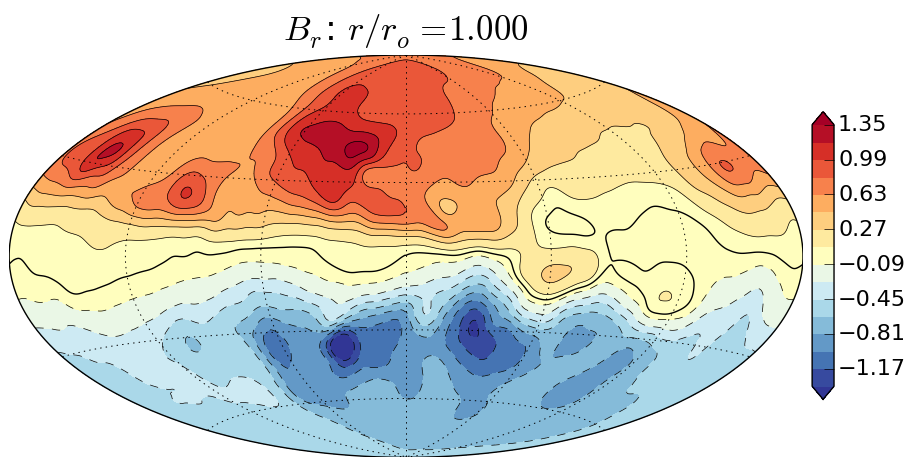} 
      \\
      \includegraphics[height=3.6cm, trim={0cm 0 0 1.3cm}, clip]{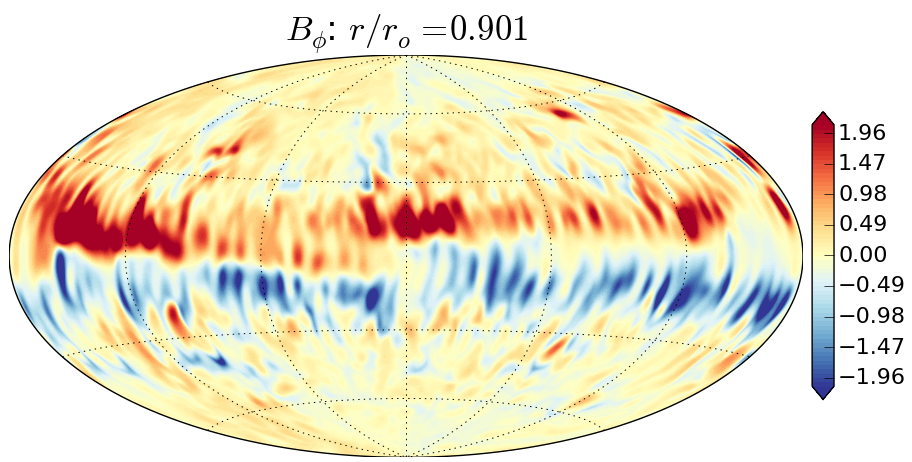}
      \includegraphics[height=3.6cm, trim={0cm 0 0 1.3cm}, clip]{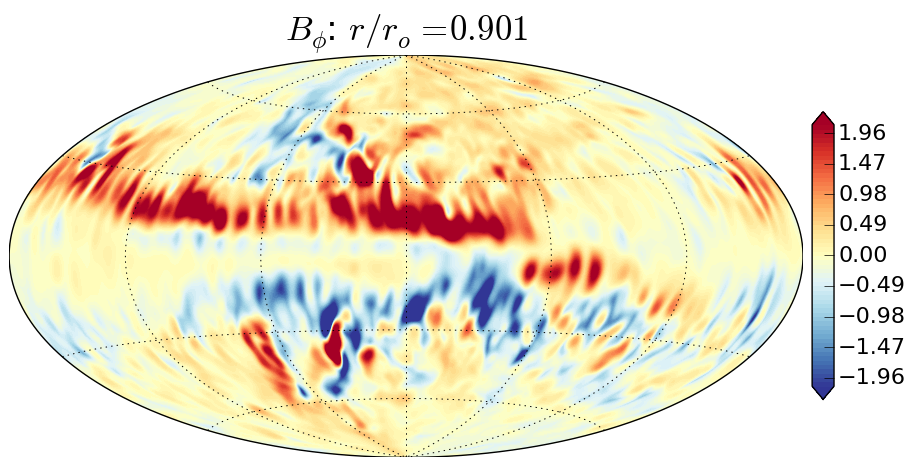}
      \\
      \includegraphics[height=3.6cm, trim={0cm 0 0 1.3cm}, clip]{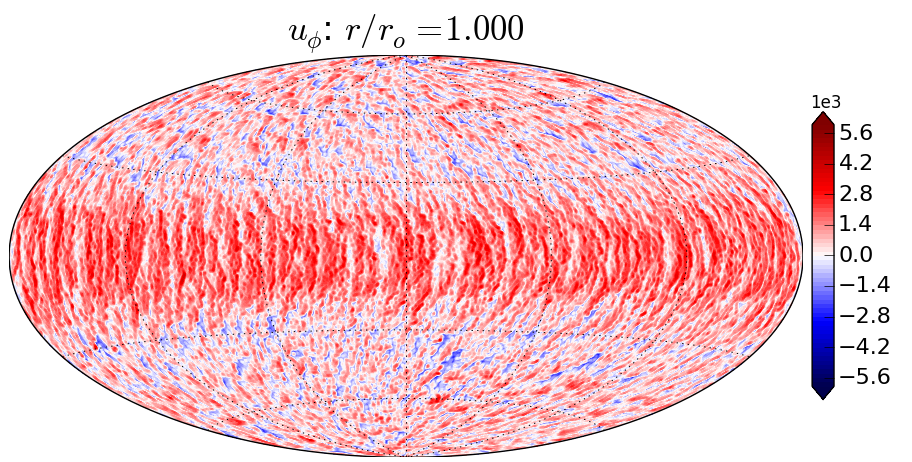}
      \includegraphics[height=3.6cm, trim={0cm 0 0 1.3cm}, clip]{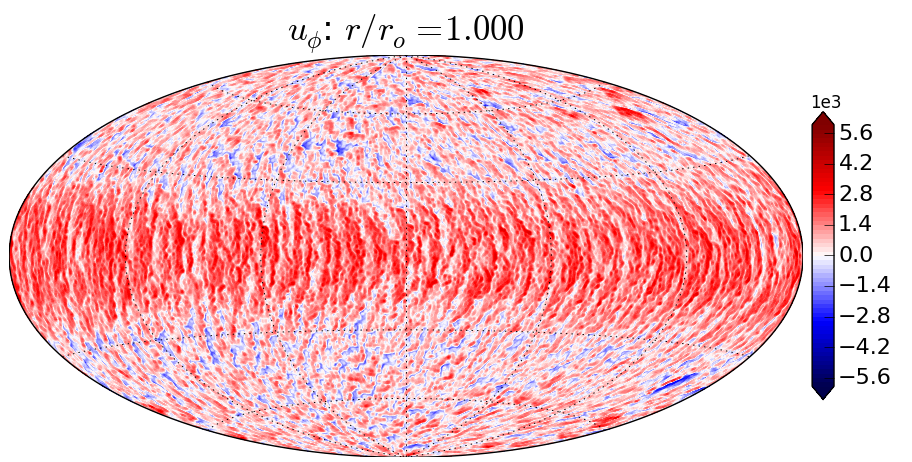}
\caption{\small{Radial magnetic field at the surface (top), azimuthal magnetic field at $\RD$ (middle) and azimuthal flow at the surface (bottom) for model $56$ snapshots with the particularly large (left) and low (right) $\mis$ values already depicted in \figref{fig:bestspecs}b. Red (blue) depicts outward (inward) radial prograde (retrograde) magnetic field, respectively.}}
\label{fig:bestbrs}
\end{figure*}

\Figref{fig:bestspecs}b illustrates the time variability in the spectrum of model $56$ which has
the lowest mean misfit \mis\ among all examined simulations.
During the model evolution, the misfit varies significantly between $0.012$ and
$0.127$ around a mean of $\mis\!=\!0.061$.
Dotted and dashed lines in \figref{fig:bestspecs}b depict the spectrum for a particularly low
value of $\mis\!=\!0.016$ and a large value of  $0.127$ respectively. \Figref{fig:bestbrs} illustrates the
respective magnetic field and zonal flow configurations.
For the small misfit snapshot on the right, quadrupole and octupole contributions agree nearly perfectly with VIP4
while the larger misfit spectrum has a not very Jupiter-like zigzag structure.
The lower panels in \figref{fig:bestbrs} depict the azimuthal flow at the outer boundary. The strong equatorial jet is clearly correlated with the banded structure, illustrating the importance of the $\Omega$-effect for the secondary dynamo operating in the transition region.

The magnetic fields show pronounced banded structures that are a result of the
secondary dynamo discussed by G14: where the equatorial zonal flow jet reaches down to sizeable electrical
conductivity values, strong azimuthal magnetic field bundles are created by the so-called $\Omega$-effect,
i.e.~shearing of radial field in azimuthal direction.
Radial flows acting on these bundles in turn create strong surface field features at low
to mid latitudes. While this process typically has a strong axisymmetric component, longitudinal
variations in background magnetic field and radial flow can also lead to significant non-axisymmetric
contributions.
The strong zigzag structure in the large \mis\ field spectrum can be traced back to
a highly equatorially antisymmetric configuration with dominant axisymmetric and
spherical harmonic order $m\!=\!2$ contributions.
The axisymmetric contribution is mostly responsible for the relatively strong optupole
field in our simulations. In VIP4 and JCF, on the other hand, the axisymmetric octupole is
surprisingly weak.
The Jupiter-like small misfit spectrum, on the other hand, is owed to a strong equatorially
antisymmetric $m\!=\!1$ band structure which boosts $\ell\!=\!2$ and, to a smaller degree,
also $\ell\!=\!4$ field contributions.

\subsection{\nad{Rescaling to Jupiter conditions}}
\label{jupiter}

\nad{
Since there are several ways that the non-dimensional results can be rescaled to physical values,
additional assumptions are required to overcome this non-uniqueness. For example, \cite{Jones14} assumes that the secular variation in the axial dipole component inferred by \cite{Ridley12} was correctly captured by his numerical dynamo model. This defines the time scale and ultimately leads to an estimated surface field that is about one order of magnitude too strong.
}

\nad{
Theoretical considerations and extensive exploration of numerical dynamo simulations for a wide variety of set-ups suggest that the mean internal field strength as well as the mean convective velocity depend on the available convective power
\citep{Christensen06,Christensen10,Aubert09,Davidson13,Yadav13a,Yadav13b}.
Support for the related scaling laws comes from the fact that they not only
successfully predict the magnetic field strength of planets but also of some rapidly rotating fully convective stars \citep{Christensen09}. Formulated in a non-dimensional framework \citep{Yadav13b}, these laws show that the Lorentz number obeys
\begin{equation}
      Lo \sim \sqrt{f_{Ohm}} \; 1.08 \,P^{*\,0.35}
      \textrm{,}
\label{eq:lorscaling}
\end{equation}
while the convective Rossby number for the non-axisymmetric flow contributions follows
\begin{equation}
      Ro_c \sim 1.65 \,P^{*\,0.42}
      \textrm{ .}
\label{eq:roscaling}
\end{equation}
These scaling laws result from fitting power law dependencies to a number of dynamo simulations that
also include anelastic models similar to the ones explored here \citep{Yadav13b} but only very few simulations
with a weakly conducting outer layer.
$P^{*}\!=\!P/(\Omega^3 d^2)$ is the dimensionless form of the convective power density $P$ (per mass) that can
be approximated from the total surface heat flux density $F_{\jupiter}$ (per surface area) via
\begin{equation}
      P = \frac{4\pi R_{\jupiter}^2 \!F_{\jupiter}}{M_{\jupiter}}\,\int \frac{\alpha \, g}{c_p} d r
         \textrm{ .}
\label{eq:power}
\end{equation}
The factor $f_{Ohm}$ in \eqnref{eq:lorscaling} is the ratio of ohmic to total dissipation.
Because of the small magnetic Prandtl number of planetary dynamo regions, ohmic dissipation clearly
dominates so that $f_{Ohm}\!\approx\!1$. In the simulations where $Pm$ is typically around order one, $f_{Ohm}$ is
smaller and typically varies around $f_{Ohm}\!\approx\!0.13$ in our simulations.
The limited spread in $Pm$ explored in numerical variations also means that the related scaling exponent is hard to constrain.
}

\nad{
\begin{figure}[h!]
\centering
      \includegraphics[width=0.5\textwidth, trim={0.4cm 0.cm -0.4cm 0.cm}, clip]{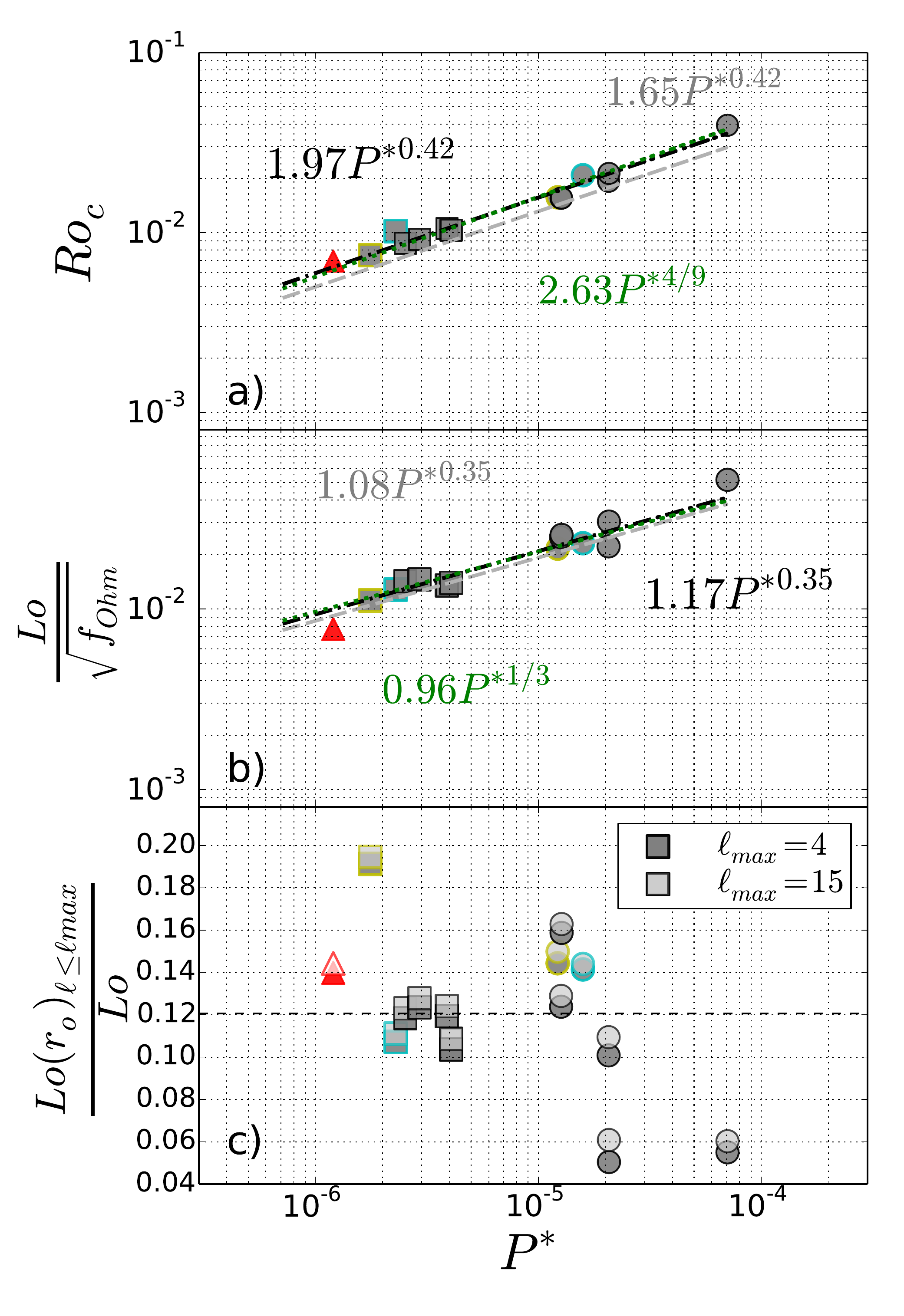}
\caption{
\small Rossby number (a) and compensated Lorentz number (b) plotted against the dimensionless power for the models listed in \tabref{tab:bestmodels}.
The grey dashed line corresponds to the scaling law (\eqnref{eq:lorscaling}, the black dot-dashed line and green dotted show polynomial fits to our best models with fixed exponents (see text for more explanation).
Panel (c) shows the ratio of surface Lorentz number $Lo_o$ to $Lo$ when considering the surface field up to spherical harmonic degree and order four (dark grey symbols) or degree and
order $15$ (light grey symbols), representing the VIP4 and the expected Juno resolutions, respectively.}
\label{fig:LoRoplots}
\end{figure}
}

\nad{
\Figref{fig:LoRoplots} demonstrates that our best dynamo simulations follow these two scaling laws (dashed grey lines).
The flow velocities are somewhat larger than suggested by \eqnref{eq:roscaling} which we attribute to the
larger flow velocities in the weakly conducting outer layer. To account for this differences, we varied the
prefactor in the scaling law and found a best fit to our simulation results for a value of $1.97$ instead of $1.65$.
An equivalent fitting for the magnetic field strength suggests a slightly larger prefactor of $1.17$ instead of $1.08$
for the Lorentz number (see black dashed line in \figref{fig:LoRoplots}).
}

\nad{
\cite{Davidson13} derives a slightly different scaling based on theoretical considerations, mostly the fact that dynamos
likely operate in a regime where Lorentz force, Coriolis force, buoyancy and pressure gradient balance in the Navier-Stokes equation.
The scalings suggested for Lorentz and total Rossby number are then $Ro\!\sim\!P^{*\,4/9}$ and $Lo\!\sim\!P^{*\,1/3}$.
Taking $Ro_c$ instead of $Ro$ to eliminate the contribution of the zonal flow, we fit the suggested scaling laws to our most Jupiter-like solutions.
This yields prefactors of $2.63$ for $Ro$ and $0.96$ for $Lo$ (see green dotted line in \eqnref{eq:roscaling}).
}

\nad{
When using the estimate of $5.5\,$W/m$^2$  for the surface heat flux density \citet{Hanel81} and the interior model by \cite{French12},
the scaling laws \eqnref{eq:lorscaling} and \eqnref{eq:roscaling}
predict values of $Ro_{c,\jupiter}\!=\!3.36\times10^{-6}$ and $Lo_{\jupiter}\!=\!1.95\times10^{-5}$, respectively.
This translates to a reasonable rms magnetic field strength of $7\,$mT and a rms convective velocity of $3\,$cm/s as already discussed by G14.
Using the adjusted prefactors translates into a $23$\% higher flow velocity and a $16$\% higher field strength which are only marginal adjustments when considering the uncertainties in the scaling procedure. The alternative scaling suggested by \cite{Davidson13} yields
an rms field strength of $12\,$mT and a rms convective velocity of $1\,$cm/s.
}

\nad{
In order to compare the rms field strength with measurements of Jupiter's surface field,
we have to establish how both are related. \Figref{fig:LoRoplots}c shows the ratio of the surface Lorentz number
\begin{equation}
      Lo_o = \left(\;\frac{E}{Pm_i} \frac{B_o^2}{\tilde{\rho}_o}\;\right)^{1/2}
      \textrm{,}
\end{equation}
filtered at $\ell_{max}\!=\!4$, to the value of $Lo\,\rho^{1/2}$ expected if the density dependence
would provide the only radial variation.
The ratio ranges between $0.05$ and $0.20$ with a value of $0.14$ for G14.
The scaling \eqnref{eq:lorscaling} would thus predict a rms surface field strength
of $0.98\,$mT for G14 which is about $40$\% higher than the VIP4 of JCF value of $0.68\,$mT.
However, the ratio of $0.10$ required to reproduce the observational field strength lies within the range of our models.
Similar inferences hold when using the \cite{Davidson13} scaling.
\Figref{fig:LoRoplots}c demonstrates that the surface field strength increases by a
maximum of $20$\% when increasing the resolution from the VIP4
values (degree and order $4$, dark grey symbols) to the expected
Juno value (degree and order $15$, light grey symbols).
}

\nad{
The scaling predictions for the rms flow velocity
between $1\,$cm/s and $10\,$cm/s agrees with similar estimates from other
authors \citep{Vasavada05,Christensen06,Jones14}.
Applying the same scaling to the zonal flow velocity, however,
yields a much too low value. \cite{Gastine12} demonstrate that the zonal
flow velocity is determined by a balance between Reynolds stress and viscous drag.
The latter is too large in the simulations in order to suppress the small
scale dynamics that cannot be resolved with the available numerical power.
Larger relative zonal flow velocities that would more clearly dominate the
convective flow would require using lower Ekman numbers which would considerably increase the numerical costs.
}

\section{Discussion and Conclusion}
\label{conclusions}

Our numerical simulations demonstrate that Jupiter's magnetic field in
the pre-Juno era can be explained by a variety of models.
Successful models require a steep density and electrical conductivity decrease
in the outer envelope similar to predictions from interior models.
While the details of the density profile hardly matter, the conductivity
profile is much more influential, mainly because it determines the
depth of the dynamo region.
As a working hypothesis we assume that dynamo action starts at the
relative radius $\RD$ where the convective magnetic Reynolds number
exceeds $50$, a critical value for self consistent numerical dynamo action
in Boussinesq models \citep{Christensen06}.
Jupiter-like models with realistic relative axial dipole, equatorial dipole,
quadrupole and octupole field contributions are those where $\RD$
corresponds to roughly $90$\% of the planetary radius.
While smaller values below $\RD\!=\!0.85$ can be excluded with some confidence
larger values however remain possible.

Assuming a rms non-axisymmetric velocity of $3\,$cm/s (G14)
and the electrical conductivity profile suggested by \citep{French12}
yields convective magnetic Reynolds numbers of more than $10^6$
at depth and $\RD\!\approx\!0.95$ for Jupiter.
Because of the generally large flow velocities the top of the dynamo region necessarily
lies in the regime where the electrical conductivity
decreases very steeply and its exact location is rather insensitive to the
flow amplitude.
Assuming, for example, a ten times larger velocity of $30\,$cm/s
would only change $\RD$ to $0.965$.

\begin{figure}[h!]
\centering
      \includegraphics[width=0.48\textwidth]{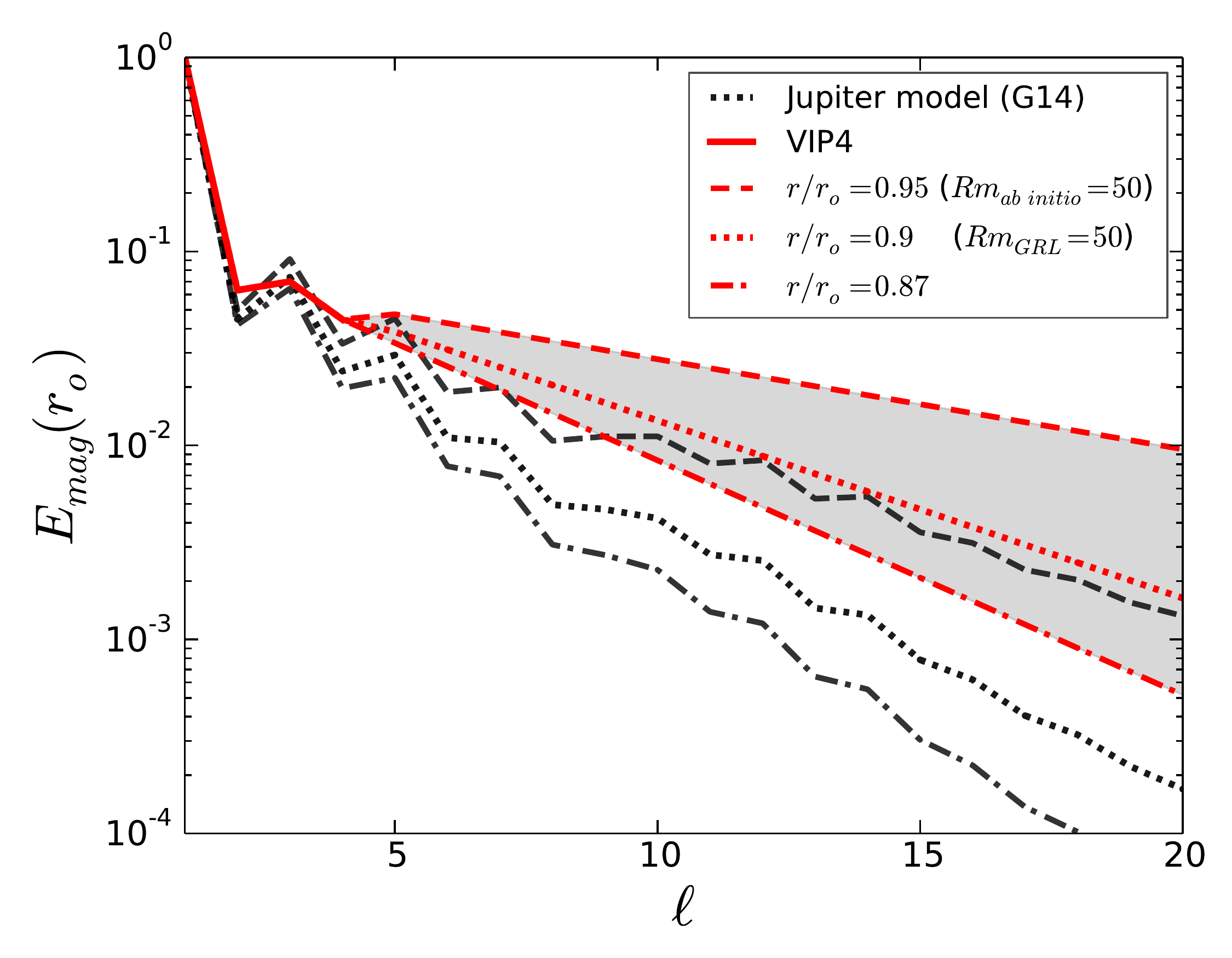}
\caption{\small Magnetic power spectra for G14 (black), and VIP4 (red). In order to anticipate the possible high degree power spectrum, VIP4 has been extended by assuming a potential field above the top of the dynamo region at $0.95 R_{\jupiter}$ (red dashed), $0.90 R_{\jupiter}$ (red dotted) and $0.87 R_{\jupiter}$ (red dashed-dotted), respectively. The field at the top of the dynamo region is supposed to be white for $\ell\!>\!4$.
The solid black line shows the G14 spectrum at $\RD\!=\!0.90$. Assuming that this would
really represent the top of the dynamo we have
calculated alternative spectra for  $0.87 R_{\jupiter}$ (black, dash-dotted)  and
$0.95 R_{\jupiter}$ (black, dashed).}
\label{fig:vip4extrap}
\end{figure}

Since we have not really explored many larger $\RD$ values in our
Jupiter-like simulations we can only estimate how such a
value would affect the numerical models.
Assuming a potential field beyond the top of the dynamo region
at $\RD\!=\!0.90$ we have calculated the surface field
spectrum for G14 at alternative values of $\RD\!=\!0.87$ and $\RD\!=\!0.95$
based on the scaling factor $(\RD/0.90)^{2\ell+4}$.
\Figref{fig:vip4extrap} demonstrates that raising the top of the dynamo region
further boosts the already large octupole contributions
in our simulations. On the other hand, the relative quadrupole and
$\ell\!=\!4$ contributions are now closer to VIP4 values and the misfit changes only slightly.
This misfit value is thus not sensitive enough to distinguish between models with $R_{50}\!>\!0.9$.
For the deeper dynamo region, the $\ell\!=\!4$ contribution becomes
much too low.
An adjustment in Rayleigh number could compensate these effects
to some degree.

Since the impact of $\RD$ grows with spherical harmonic degree
$\ell$, the high resolution Juno results will allow to
constrain the depth of the dynamo region much better than VIP4.
Assuming a white spectrum for $\ell\!>\!4$ at $\RD$ leads to the
predicted spectra shown in \figref{fig:vip4extrap}.
At $\ell\!=\!20$ the differences between the $\RD\!=\!0.90$ and $\RD\!=\!0.95$
spectra amount to more than an order of magnitude.
We note that the respective G14 spectra show a somewhat different
tilt which also depends on the $\ell$ range. This suggests that
the G14 spectrum is not really white at the top of the dynamo region,
a topic that deserved to be addressed in more detail.

The secondary dynamo due to the equatorial jet can produce strong magnetic
bands at low to mid-latitudes on both sides of the equator.
The magnetic field agrees more convincingly with the VIP4 model
\citep{Connerney98} when these features are concentrated in one hemisphere.
Juno's measurement should be able to resolve these features that
allow to constrain the deep zonal jet dynamics.

The unrealistically small magnetic Reynolds numbers in the simulations
are unsatisfactory.
While values of up to $Rm\!=\!10^6$ are expected for Jupiter's interior,
our numerical models only reach $1000$.

The simulations indicate that a larger $Rm$ tends to boost
higher field harmonics beyond $\ell\!=\!3$ which suggest
another feature that should be constrained by the Juno mission.

Larger $Rm$ values could be reached by either increasing the
magnetic Prandtl number $Pm$ or the Rayleigh number.
Planetary dynamos are characterized by magnetic Prandtl numbers
much smaller than unity which means that Ohmic diffusion clearly dominates
viscous diffusion. Since $Pm_i$ is already larger than unity in our simulations,
increasing it even further seems like the wrong way to go.
Increasing the Rayleigh number, on the other hand, always bears the
danger of venturing into the multipolar regime unless the Ekman number
is decreased accordingly \citep{Christensen06}.
Since both larger $Ra$ and smaller $E$ lead to smaller spatial scales
and shorter time steps, the numerical costs prevented us from
following this approach further.

Our attempts to explore the dependence on the Ekman number and
on the heating mode remain inconclusive.
Lower Ekman numbers indeed seem to promote
large scale field production in Boussinesq dynamos because of the stronger Coriolis force
that helps to organize the flow \citep{Christensen06}.
However, our database is still too limited to confirm a similar effect in
anelastic simulations.
All our nine
fully internally heated simulations ended up as multipolar dynamos.
This issue was already reported by \cite{Jones14} who nevertheless found a few Jupiter-like internally heated cases. A special combination of a low Ekman number, a low Prandtl number and a stronger concentration of heat sources at depth may have helped in his model. The difficulties in finding dipole-dominated solutions with internal heating seem remarkable since this is the more realistic driving scenario for Jupiter's internal dynamo. We can only speculate that the problem may vanish at more realistic parameters, for example at lower Ekman
numbers.

\cite{Sreenivasan06} and \cite{Simitev09} show that at low Rayleigh numbers large $Pr$ Boussinesq dynamos are dipole-dominated
while small $Pr$ dynamos are multipolar. Our anelastic simulations seem to confirm this but also indicate that the small $Pr$ dynamos become dipole-dominated at larger Rayleigh numbers.
Decreasing the Prandtl number below the $Pr\!=\!0.1$ value
may thus help to reconcile dipole-dominated dynamo with internal heating.

\section*{Acknowledgements}

All the computations have been carried out in the GWDG computer facilities in
G\"ottingen.

This work was partially supported by the Special Priority Program 1488
(PlanetMag, {http://www.planetmag.de}) of the German Science Foundation.

We would like to thank the reviewers for the very helpful comments. We would also like to thank Chris Jones and Wieland Dietrich for kindly compiling and providing the data added after the revision to make our parameter study much more complete.


\bibliographystyle{elsarticle-harv}
\bibliography{biblio}

{\footnotesize

\onecolumn
\begin{center}

\begingroup
\setlength{\tabcolsep}{4.8pt}

\begin{longtable}{ccccccccccccccccccc}

\caption{Summary of the time-averaged results. The grey-coloured rows correspond to dipole-dominated cases. In the column `$BC$', the cases with octupole entropy at both boundaries are represented by $S\!S$ and the cases with fixed flux at both boundaries are represented by $F\!F$. The supercriticality of each model in column $Ra/Ra_{cr}$ is calculated using the values of $Ra_{cr}$ listed in Tab.~\ref{Tab1}.
\small{
\label{TabRes}}
}\\

\hline
\vspace{-6.5pt}
 & & & \\

Model & $N_{\rho}$ & $\begin{matrix}E\\10^{-5}\end{matrix}$ & $\dfrac{Ra}{Ra_{cr}}$ & $Pr$ & $Pm_i$ & $Pm_V$ & $a/\sigma_m/\chi_m$ & $H$ & $BC$ & $\begin{matrix}f_{dip}\\(\ell_{max}\!=\!4)\end{matrix}$ & $\begin{matrix}\theta_{dip}\\(^\circ)\end{matrix}$ & $\dfrac{ B_{\,\ell=2} }{ B_{\,\ell\le4} }$ & $\dfrac{ B_{\,\ell=3} }{ B_{\,\ell\le4} }$ & $Ro_{zon}$ & $Rm_c$ & $\Lambda$ & $\begin{matrix}r_{50}\\(\%)\end{matrix}$ & $\mis$ \\
\hline
\vspace{-9.5pt}
 & & & \\
\hline
\endfirsthead

\hline
Model & $N_{\rho}$ & $\begin{matrix}E\\10^{-5}\end{matrix}$ & $\dfrac{Ra}{Ra_{cr}}$ & $Pr$ & $Pm_i$ & $Pm_V$ & $a/\sigma_m/\chi_m$ & $H$ & $BC$ & $\begin{matrix}f_{dip}\\(\ell_{max}\!=\!4)\end{matrix}$ & $\begin{matrix}\theta_{dip}\\(^\circ)\end{matrix}$ & $\dfrac{ B_{\,\ell=2} }{ B_{\,\ell\le4} }$ & $\dfrac{ B_{\,\ell=3} }{ B_{\,\ell\le4} }$ & $Ro_{zon}$ & $Rm_c$ & $\Lambda$ & $\begin{matrix}r_{50}\\(\%)\end{matrix}$ & $\mis$ \\
\hline
\vspace{-9.5pt}
 & & & \\
\hline
\endhead

\hline
\multicolumn{15}{r}{{Continued on next page}} \\
\hline
\endfoot
\endlastfoot

\vspace{-8pt}
 & & & \\

 \rowcolor{lightgray} $1$ & $4$ & $10$ & $5.5$ & $1.0$ & $4.0$ & $2.05$ & $9./0.500/80$ & $1.0$ & $S\!S$ & $8.22\!\times\!10^{-1}$ & $5.8$ & $0.199$ & $0.319$ & $1.15\!\times\!10^{-2}$ & $156$ & $2.04$ & $92.6$ & $0.067$ \\ 
 \rowcolor{lightgray} $2$ & $4$ & $3$ & $5.6$ & $1.0$ & $2.0$ & $1.03$ & $9./0.500/80$ & $1.0$ & $S\!S$ & $9.00\!\times\!10^{-1}$ & $4.0$ & $0.165$ & $0.227$ & $6.65\!\times\!10^{-3}$ & $115$ & $0.99$ & $89.1$ & $0.065$ \\ 

 \hline

\rowcolor{lightgray} $3$ & $5$ & $10$ & $7.4$ & $1.0$ & $2.0$ & $1.02$ & $9./0.500/80$ & $1.0$ & $S\!S$ & $9.32\!\times\!10^{-1}$ & $1.4$ & $0.077$ & $0.229$ & $1.41\!\times\!10^{-2}$ & $63$ & $0.63$ & $83.3$ & $0.106$ \\ 
\rowcolor{lightgray} $4$ & $5$ & $10$ & $7.4$ & $1.0$ & $2.0$ & $1.27$ & $13./0.200/90$ & $1.0$ & $S\!S$ & $8.86\!\times\!10^{-1}$ & $2.5$ & $0.102$ & $0.300$ & $1.47\!\times\!10^{-2}$ & $80$ & $0.62$ & $89.2$ & $0.096$ \\ 
\rowcolor{lightgray} $5$ & $5$ & $10$ & $7.4$ & $1.0$ & $2.0$ & $0.99$ & $17./0.500/80$ & $1.0$ & $S\!S$ & $9.47\!\times\!10^{-1}$ & $1.1$ & $0.058$ & $0.207$ & $1.92\!\times\!10^{-2}$ & $51$ & $0.36$ & $80.0$ & $0.117$ \\ 
\rowcolor{lightgray} $6$ & $5$ & $10$ & $7.4$ & $1.0$ & $2.0$ & $0.98$ & $25./0.500/80$ & $1.0$ & $S\!S$ & $9.60\!\times\!10^{-1}$ & $1.0$ & $0.048$ & $0.178$ & $1.98\!\times\!10^{-2}$ & $48$ & $0.30$ & $78.9$ & $0.126$ \\ 
\rowcolor{lightgray} $7$ & $5$ & $10$ & $7.4$ & $1.0$ & $2.0$ & $1.00$ & $40./0.500/80$ & $1.0$ & $S\!S$ & $9.62\!\times\!10^{-1}$ & $1.3$ & $0.053$ & $0.163$ & $2.06\!\times\!10^{-2}$ & $48$ & $0.29$ & $80.0$ & $0.125$ \\ 
\rowcolor{lightgray} $8$ & $5$ & $10$ & $9.3$ & $1.0$ & $2.0$ & $1.27$ & $13./0.200/90$ & $1.0$ & $S\!S$ & $7.62\!\times\!10^{-1}$ & $9.9$ & $0.258$ & $0.334$ & $1.53\!\times\!10^{-2}$ & $114$ & $1.08$ & $90.1$ & $0.070$ \\ 
\rowcolor{lightgray} $9$ & $5$ & $10$ & $9.3$ & $1.0$ & $2.5$ & $0.60$ & $1./0.030/90$ & $1.0$ & $S\!S$ & $9.85\!\times\!10^{-1}$ & $1.1$ & $0.052$ & $0.083$ & $2.46\!\times\!10^{-2}$ & $39$ & $0.48$ & $74.9$ & $0.144$ \\ 
 $10$ & $5$ & $10$ & $9.3$ & $1.0$ & $2.0$ & $1.03$ & $9./0.500/80$ & $1.0$ & $S\!S$ & $6.23\!\times\!10^{-1}$ & $19.6$ & $0.408$ & $0.289$ & $2.16\!\times\!10^{-2}$ & $87$ & $0.39$ & $86.5$ & $0.138$ \\ 
\rowcolor{lightgray} $11$ & $5$ & $10$ & $9.3$ & 1.0 & $2.0$ & $1.03$ & $9./0.500/80$ & $1.0$ & $S\!S$ & $8.65\!\times\!10^{-1}$ & $3.7$ & $0.165$ & $0.283$ & $1.55\!\times\!10^{-2}$ & $81$ & $1.00$ & $85.8$ & $0.067$ \\ 
 $12$ & $5$ & $10$ & $11.2$ & $1.0$ & $2.0$ & $0.69$ & $9./0.500/70$ & $1.0$ & $S\!S$ & $3.57\!\times\!10^{-1}$ & $43.7$ & $0.461$ & $0.509$ & $3.92\!\times\!10^{-2}$ & $51$ & $0.21$ & $72.4$ & $0.283$ \\ 
\rowcolor{lightgray} $13$ & $5$ & $3$ & $9.2$ & $1.0$ & $1.5$ & $0.96$ & $13./0.200/90$ & $1.0$ & $S\!S$ & $8.28\!\times\!10^{-1}$ & $7.7$ & $0.185$ & $0.316$ & $1.05\!\times\!10^{-2}$ & $129$ & $0.64$ & $90.4$ & $0.061$ \\ 
\rowcolor{lightgray} $14$ & $5$ & $3$ & $16.1$ & $1.0$ & $1.0$ & $0.64$ & $13./0.200/90$ & $1.0$ & $S\!S$ & $7.64\!\times\!10^{-1}$ & $10.8$ & $0.193$ & $0.376$ & $1.42\!\times\!10^{-2}$ & $165$ & $1.12$ & $90.6$ & $0.086$ \\ 
\rowcolor{lightgray} $15$ & $5$ & $1$ & $10.4$ & $1.0$ & $0.6$ & $0.38$ & $13./0.200/90$ & $1.0$ & $S\!S$ & $7.91\!\times\!10^{-1}$ & $9.4$ & $0.126$ & $0.397$ & $1.11\!\times\!10^{-2}$ & $140$ & $0.33$ & $90.5$ & $0.099$ \\ 

\hline

 $16$ & $\jupiter$ & $10$ & $7.8$ & $0.1$ & $2.0$ & $1.27$ & $13./0.200/90$ & $1.0$ & $S\!S$ & $3.83\!\times\!10^{-2}$ & $24.2$ & $0.496$ & $0.631$ & $2.56\!\times\!10^{-2}$ & $158$ & $0.12$ & $89.5$ & $0.430$ \\ 
 $17$ & $\jupiter$ & $10$ & $8.2$ & $0.1$ & $2.0$ & $1.27$ & $13./0.200/90$ & $1.0$ & $S\!S$ & $3.32\!\times\!10^{-2}$ & $37.0$ & $0.476$ & $0.625$ & $2.75\!\times\!10^{-2}$ & $162$ & $0.20$ & $89.4$ & $0.432$ \\ 
 $18$ & $\jupiter$ & $10$ & $8.4$ & $0.1$ & $2.0$ & $1.27$ & $13./0.200/90$ & $1.0$ & $S\!S$ & $5.57\!\times\!10^{-3}$ & $73.8$ & $0.618$ & $0.238$ & $2.09\!\times\!10^{-2}$ & $207$ & $4.63$ & $89.5$ & $0.476$ \\ 
\rowcolor{lightgray} $19$ & $\jupiter$ & $10$ & $9.7$ & $0.1$ & $2.0$ & $1.27$ & $13./0.200/90$ & $1.0$ & $S\!S$ & $8.98\!\times\!10^{-1}$ & $3.4$ & $0.139$ & $0.218$ & $2.78\!\times\!10^{-2}$ & $206$ & $4.85$ & $89.4$ & $0.087$ \\ 
\rowcolor{lightgray} $20$ & $\jupiter$ & $10$ & $11.7$ & $0.1$ & $2.0$ & $1.27$ & $13./0.200/90$ & $1.0$ & $S\!S$ & $8.75\!\times\!10^{-1}$ & $4.6$ & $0.170$ & $0.221$ & $3.09\!\times\!10^{-2}$ & $264$ & $7.49$ & $90.6$ & $0.064$ \\ 

 $21$ & $\jupiter$ & $10$ & $9.4$ & $0.1$ & $4.0$ & $2.53$ & $13./0.200/90$ & $0.0$ & $F\!F$ & $1.83\!\times\!10^{-3}$ & $84.1$ & $0.602$ & $0.216$ & $1.73\!\times\!10^{-2}$ & $247$ & $3.75$ & $90.4$ & $0.483$ \\ 
 $22$ & $\jupiter$ & $10$ & $9.4$ & $0.1$ & $8.0$ & $5.05$ & $13./0.200/90$ & $0.0$ & $F\!F$ & $5.29\!\times\!10^{-1}$ & $85.9$ & $0.216$ & $0.517$ & $1.94\!\times\!10^{-2}$ & $419$ & $4.70$ & $91.4$ & $0.199$ \\ 
 $23$ & $\jupiter$ & $10$ & $14.4$ & $0.1$ & $1.0$ & $0.63$ & $13./0.200/90$ & $0.0$ & $F\!F$ & $1.70\!\times\!10^{-3}$ & $88.5$ & $0.750$ & $0.145$ & $2.33\!\times\!10^{-2}$ & $97$ & $1.41$ & $89.5$ & $0.515$ \\ 
 $24$ & $\jupiter$ & $10$ & $14.4$ & $0.1$ & $2.0$ & $1.27$ & $13./0.200/90$ & $0.0$ & $F\!F$ & $2.96\!\times\!10^{-3}$ & $87.5$ & $0.610$ & $0.231$ & $2.01\!\times\!10^{-2}$ & $189$ & $3.04$ & $90.6$ & $0.476$ \\ 
 $25$ & $\jupiter$ & $10$ & $18.1$ & $0.1$ & $2.0$ & $1.27$ & $13./0.200/90$ & $0.0$ & $F\!F$ & $1.25\!\times\!10^{-1}$ & $87.0$ & $0.482$ & $0.416$ & $3.29\!\times\!10^{-2}$ & $211$ & $2.49$ & $90.6$ & $0.392$ \\ 

\rowcolor{lightgray} $26$ & $\jupiter$ & $10$ & $5.7$ & $1.0$ & $2.0$ & $1.27$ & $13./0.200/90$ & $1.0$ & $S\!S$ & $8.38\!\times\!10^{-1}$ & $3.3$ & $0.143$ & $0.331$ & $1.03\!\times\!10^{-2}$ & $83$ & $0.87$ & $88.3$ & $0.091$ \\ 
 $27$ & $\jupiter$ & $10$ & $5.7$ & $1.0$ & $2.0$ & $0.98$ & $25./0.500/80$ & $1.0$ & $S\!S$ & $1.74\!\times\!10^{-1}$ & $79.5$ & $0.646$ & $0.441$ & $1.63\!\times\!10^{-2}$ & $54$ & $0.20$ & $79.0$ & $0.387$ \\ 
\rowcolor{lightgray} $28$ & $\jupiter$ & $10$ & $5.7$ & $1.0$ & $2.0$ & $1.20$ & $36./0.500/85$ & $1.0$ & $S\!S$ & $8.78\!\times\!10^{-1}$ & $1.8$ & $0.102$ & $0.294$ & $1.20\!\times\!10^{-2}$ & $74$ & $0.77$ & $85.2$ & $0.105$ \\ 
 $29$ & $\jupiter$ & $10$ & $5.7$ & $1.0$ & $3.0$ & $1.40$ & $6./0.400/80$ & $1.0$ & $S\!S$ & $3.21\!\times\!10^{-1}$ & $87.4$ & $0.498$ & $0.435$ & $1.41\!\times\!10^{-2}$ & $84$ & $0.35$ & $83.2$ & $0.303$ \\ 
 \rowcolor{lightgray} $30$ & $\jupiter$ & $10$ & $5.7$ & $1.0$ & $3.0$ & $1.40$ & $6./0.400/80$ & $1.0$ & $S\!S$ & $9.33\!\times\!10^{-1}$ & $2.5$ & $0.098$ & $0.201$ & $1.05\!\times\!10^{-2}$ & $81$ & $1.16$ & $83.6$ & $0.098$ \\ 
\rowcolor{lightgray} $31$ & $\jupiter$ & $10$ & $5.7$ & $1.0$ & $2.0$ & $1.03$ & $9./0.500/80$ & $1.0$ & $S\!S$ & $9.44\!\times\!10^{-1}$ & $2.5$ & $0.092$ & $0.189$ & $1.06\!\times\!10^{-2}$ & $64$ & $0.84$ & $82.0$ & $0.098$ \\ 
 $32$ & $\jupiter$ & $10$ & $6.1$ & $1.0$ & $1.0$ & $0.63$ & $13./0.200/90$ & $1.0$ & $S\!S$ & $3.93\!\times\!10^{-1}$ & $78.2$ & $0.560$ & $0.309$ & $1.68\!\times\!10^{-2}$ & $41$ & $0.16$ & $86.2$ & $0.273$ \\ 
\rowcolor{lightgray} $33$ & $\jupiter$ & $10$ & $6.1$ & $1.0$ & $2.0$ & $1.27$ & $13./0.200/90$ & $1.0$ & $S\!S$ & $8.34\!\times\!10^{-1}$ & $5.2$ & $0.157$ & $0.328$ & $1.03\!\times\!10^{-2}$ & $91$ & $0.97$ & $89.6$ & $0.077$ \\ 
\rowcolor{lightgray} $34$ & $\jupiter$ & $10$ & $6.1$ & $1.0$ & $3.0$ & $1.92$ & $13./0.200/90$ & $1.0$ & $S\!S$ & $7.47\!\times\!10^{-1}$ & $9.1$ & $0.246$ & $0.348$ & $9.84\!\times\!10^{-3}$ & $139$ & $1.40$ & $89.8$ & $0.089$ \\ 
 $35$ & $\jupiter$ & $10$ & $6.1$ & $1.0$ & $3.0$ & $0.71$ & $1./0.030/90$ & $1.0$ & $S\!S$ & $1.75\!\times\!10^{-1}$ & $67.7$ & $0.723$ & $0.307$ & $1.95\!\times\!10^{-2}$ & $38$ & $0.11$ & $71.0$ & $0.402$ \\ 
 $36$ & $\jupiter$ & $10$ & $6.1$ & $1.0$ & $2.0$ & $0.89$ & $3./0.250/85$ & $1.0$ & $S\!S$ & $3.37\!\times\!10^{-1}$ & $57.1$ & $0.509$ & $0.457$ & $1.62\!\times\!10^{-2}$ & $60$ & $0.24$ & $82.9$ & $0.291$ \\ 
\rowcolor{lightgray} $37$ & $\jupiter$ & $10$ & $6.1$ & $1.0$ & $3.0$ & $1.33$ & $3./0.250/85$ & $1.0$ & $S\!S$ & $9.31\!\times\!10^{-1}$ & $2.8$ & $0.112$ & $0.195$ & $1.02\!\times\!10^{-2}$ & $90$ & $1.31$ & $87.8$ & $0.092$ \\ 
\rowcolor{lightgray} $38$ & $\jupiter$ & $10$ & $6.1$ & $1.0$ & $2.0$ & $0.99$ & $4./0.110/90$ & $1.0$ & $S\!S$ & $8.94\!\times\!10^{-1}$ & $3.4$ & $0.106$ & $0.273$ & $1.18\!\times\!10^{-2}$ & $65$ & $0.84$ & $86.8$ & $0.088$ \\ 
\rowcolor{lightgray} $39$ & $\jupiter$ & $10$ & $6.1$ & $1.0$ & $2.0$ & $1.10$ & $5./0.200/90$ & $1.0$ & $S\!S$ & $8.75\!\times\!10^{-1}$ & $4.8$ & $0.129$ & $0.287$ & $1.00\!\times\!10^{-2}$ & $77$ & $0.94$ & $88.5$ & $0.078$ \\ 
\rowcolor{lightgray} $40$ & $\jupiter$ & $10$ & $6.1$ & $1.0$ & $2.0$ & $1.03$ & $9./0.500/80$ & $1.0$ & $S\!S$ & $9.53\!\times\!10^{-1}$ & $2.6$ & $0.089$ & $0.155$ & $1.16\!\times\!10^{-2}$ & $69$ & $0.94$ & $82.5$ & $0.109$ \\ 
 $41$ & $\jupiter$ & $10$ & $6.7$ & $1.0$ & $2.0$ & $1.02$ & $9./0.500/80$ & $1.0$ & $S\!S$ & $3.97\!\times\!10^{-1}$ & $82.9$ & $0.480$ & $0.392$ & $1.69\!\times\!10^{-2}$ & $79$ & $0.43$ & $83.4$ & $0.260$ \\ 
 $42$ & $\jupiter$ & $10$ & $7.4$ & $1.0$ & $1.0$ & $0.63$ & $13./0.200/90$ & $1.0$ & $S\!S$ & $3.39\!\times\!10^{-1}$ & $55.5$ & $0.536$ & $0.363$ & $2.00\!\times\!10^{-2}$ & $55$ & $0.32$ & $87.4$ & $0.304$ \\ 
\rowcolor{lightgray} $43$ & $\jupiter$ & $10$ & $7.4$ & $1.0$ & $2.0$ & $1.27$ & $13./0.200/90$ & $1.0$ & $S\!S$ & $8.08\!\times\!10^{-1}$ & $8.3$ & $0.191$ & $0.306$ & $1.17\!\times\!10^{-2}$ & $119$ & $1.45$ & $89.7$ & $0.074$ \\ 
\rowcolor{lightgray} $44$ & $\jupiter$ & $10$ & $7.4$ & $1.0$ & $10.$ & $6.39$ & $13./0.200/90$ & $1.0$ & $S\!S$ & $7.57\!\times\!10^{-1}$ & $7.2$ & $0.222$ & $0.301$ & $3.65\!\times\!10^{-3}$ & $565$ & $7.29$ & $92.4$ & $0.067$ \\ 
 $45$ & $\jupiter$ & $10$ & $7.4$ & $1.0$ & $4.0$ & $1.00$ & $1.1/0.020/90$ & $1.0$ & $S\!S$ & $3.65\!\times\!10^{-1}$ & $80.5$ & $0.497$ & $0.372$ & $2.23\!\times\!10^{-2}$ & $76$ & $0.42$ & $83.5$ & $0.285$ \\ 
 $46$ & $\jupiter$ & $10$ & $7.4$ & $1.0$ & $8.0$ & $2.00$ & $1.1/0.020/90$ & $1.0$ & $S\!S$ & $2.63\!\times\!10^{-1}$ & $71.2$ & $0.490$ & $0.441$ & $1.80\!\times\!10^{-2}$ & $159$ & $1.06$ & $87.6$ & $0.311$ \\ 
 $47$ & $\jupiter$ & $10$ & $7.4$ & $1.0$ & $10.$ & $2.46$ & $1.1/0.020/90$ & $1.0$ & $S\!S$ & $1.51\!\times\!10^{-1}$ & $69.4$ & $0.553$ & $0.477$ & $1.66\!\times\!10^{-2}$ & $188$ & $1.28$ & $87.4$ & $0.394$ \\ 
 $48$ & $\jupiter$ & $10$ & $8.4$ & $1.0$ & $4.0$ & $1.00$ & $1.1/0.020/90$ & $1.0$ & $S\!S$ & $5.00\!\times\!10^{-1}$ & $73.0$ & $0.404$ & $0.376$ & $2.50\!\times\!10^{-2}$ & $90$ & $0.61$ & $85.0$ & $0.210$ \\ 
 $49$ & $\jupiter$ & $10$ & $8.4$ & $1.0$ & $2.0$ & $1.27$ & $13./0.200/90$ & $1.0$ & $S\!S$ & $2.49\!\times\!10^{-1}$ & $69.3$ & $0.488$ & $0.462$ & $1.54\!\times\!10^{-2}$ & $140$ & $1.13$ & $89.7$ & $0.320$ \\ 

 $50$ & $\jupiter$ & $10$ & $5.1$ & $1.0$ & $8.0$ & $5.11$ & $13./0.200/90$ & $0.8$ & $S\!S$ & $4.65\!\times\!10^{-1}$ & $84.3$ & $0.401$ & $0.374$ & $7.46\!\times\!10^{-3}$ & $234$ & $1.39$ & $91.1$ & $0.219$ \\ 
 $51$ & $\jupiter$ & $10$ & $6.2$ & $1.0$ & $4.0$ & $2.53$ & $13./0.200/90$ & $0.9$ & $S\!S$ & $3.28\!\times\!10^{-1}$ & $54.7$ & $0.464$ & $0.434$ & $1.04\!\times\!10^{-2}$ & $161$ & $0.97$ & $89.4$ & $0.277$ \\ 
 $52$ & $\jupiter$ & $10$ & $6.2$ & $1.0$ & $6.0$ & $3.77$ & $13./0.200/90$ & $0.8$ & $S\!S$ & $1.52\!\times\!10^{-1}$ & $85.2$ & $0.525$ & $0.510$ & $8.66\!\times\!10^{-3}$ & $242$ & $1.54$ & $90.1$ & $0.370$ \\ 
 $53$ & $\jupiter$ & $10$ & $9.9$ & $1.0$ & $8.0$ & $5.11$ & $13./0.200/90$ & $0.0$ & $F\!F$ & $1.80\!\times\!10^{-1}$ & $44.9$ & $0.489$ & $0.454$ & $1.44\!\times\!10^{-2}$ & $153$ & $0.03$ & $91.1$ & $0.337$ \\ 
 $54$ & $\jupiter$ & $10$ & $11.8$ & $1.0$ & $8.0$ & $5.11$ & $13./0.200/90$ & $0.0$ & $F\!F$ & $8.67\!\times\!10^{-2}$ & $69.5$ & $0.491$ & $0.528$ & $1.65\!\times\!10^{-2}$ & $189$ & $0.18$ & $91.1$ & $0.409$ \\ 

 \rowcolor{lightgray} $55$ & $\jupiter$ & $3$ & $7.4$ & $1.0$ & $1.2$ & $0.77$ & $13./0.200/90$ & $1.0$ & $S\!S$ & $8.70\!\times\!10^{-1}$ & $5.6$ & $0.118$ & $0.306$ & $7.79\!\times\!10^{-3}$ & $98$ & $0.57$ & $89.8$ & $0.076$ \\ 
 \rowcolor{lightgray} $56$ & $\jupiter$ & $3$ & $7.4$ & $1.0$ & $1.5$ & $0.96$ & $13./0.200/90$ & $1.0$ & $S\!S$ & $8.19\!\times\!10^{-1}$ & $7.2$ & $0.221$ & $0.291$ & $7.76\!\times\!10^{-3}$ & $124$ & $0.69$ & $89.9$ & $0.061$ \\ 
 \rowcolor{lightgray} $57$ & $\jupiter$ & $3$ & $7.9$ & $1.0$ & $1.5$ & $0.96$ & $13./0.200/90$ & $1.0$ & $S\!S$ & $8.36\!\times\!10^{-1}$ & $5.9$ & $0.193$ & $0.282$ & $8.17\!\times\!10^{-3}$ & $131$ & $0.73$ & $90.1$ & $0.078$ \\ 
 \rowcolor{lightgray} $58$ & $\jupiter$ & $3$ & $8.7$ & $1.0$ & $1.2$ & $0.77$ & $13./0.200/90$ & $1.0$ & $S\!S$ & $8.06\!\times\!10^{-1}$ & $6.9$ & $0.209$ & $0.324$ & $9.11\!\times\!10^{-3}$ & $122$ & $0.67$ & $89.7$ & $0.070$ \\ 
 \rowcolor{lightgray} $59$ & $\jupiter$ & $3$ & $8.7$ & $1.0$ & $1.5$ & $0.96$ & $13./0.200/90$ & $1.0$ & $S\!S$ & $8.00\!\times\!10^{-1}$ & $5.7$ & $0.237$ & $0.312$ & $8.81\!\times\!10^{-3}$ & $152$ & $0.89$ & $90.4$ & $0.072$ \\ 
 \rowcolor{lightgray} $60$ & $\jupiter$ & $3$ & $9.6$ & $1.0$ & $3.0$ & $0.74$ & $1.1/0.015/90$ & $1.0$ & $S\!S$ & $9.29\!\times\!10^{-1}$ & $3.4$ & $0.104$ & $0.207$ & $1.43\!\times\!10^{-2}$ & $101$ & $0.98$ & $85.7$ & $0.095$ \\ 
 $61$ & $\jupiter$ & $3$ & $12.2$ & $1.0$ & $3.0$ & $0.74$ & $1.1/0.015/90$ & $1.0$ & $S\!S$ & $4.09\!\times\!10^{-1}$ & $65.6$ & $0.472$ & $0.353$ & $1.84\!\times\!10^{-2}$ & $148$ & $0.66$ & $87.0$ & $0.250$ \\ 

 \rowcolor{lightgray} $62$ & $\jupiter$ & $3$ & $10.2$ & $1.0$ & $3.0$ & $0.74$ & $1.1/0.015/90$ & $0.9$ & $S\!S$ & $9.34\!\times\!10^{-1}$ & $3.5$ & $0.111$ & $0.189$ & $1.47\!\times\!10^{-2}$ & $99$ & $0.96$ & $85.7$ & $0.092$ \\
$63$ & $\jupiter$ & $3$ & $22.0$ & $1.0$ & $4.0$ & $0.99$ & $1.1/0.015/90$ & $0.0$ & $F\!F$ & $6.26\!\times\!10^{-1}$ & $41.0$ & $0.339$ & $0.373$ & $1.71\!\times\!10^{-2}$ & $59$ & $0.06$ & $84.3$ & $0.142$ \\ 
 $64$ & $\jupiter$ & $3$ & $23.7$ & $1.0$ & $4.0$ & $0.99$ & $1.1/0.015/90$ & $0.0$ & $F\!F$ & $2.79\!\times\!10^{-1}$ & $85.6$ & $0.547$ & $0.427$ & $1.74\!\times\!10^{-2}$ & $64$ & $0.09$ & $85.2$ & $0.314$ \\ 

 $65$ & $\jupiter$ & $1$ & $12.3$ & $1.0$ & $1.5$ & $0.37$ & $1.1/0.015/90$ & $1.0$ & $S\!S$ & $3.15\!\times\!10^{-1}$ & $89.8$ & $0.546$ & $0.349$ & $1.43\!\times\!10^{-2}$ & $102$ & $0.16$ & $85.2$ & $0.319$ \\ 

\hline
\vspace{-8pt}
 & & & \\

 G14 & $\jupiter$ & $1$ & $12.3$ & $1.0$ & $0.6$ & $0.37$ & $13./0.200/90$ & $1.0$ & $S\!S$ & $8.46\!\times\!10^{-1}$ & $7.0$ & $0.208$ & $0.240$ & $9.54\!\times\!10^{-3}$ & $124$ & $-$ & $89.5$ & $0.063$ \\ 
\hline
\vspace{-18pt}
 & & & \\
 \\

\hline

\end{longtable}
\endgroup

\end{center}
\twocolumn

}


\end{document}